\documentclass[12pt]{article}
\usepackage{amsmath,amsthm,amsbsy,amsfonts,comment,booktabs}
\usepackage{bm,graphicx,psfrag,epsf,booktabs,bbm,multirow}
\usepackage{enumerate}
\usepackage[toc,page]{appendix}
\usepackage[round, authoryear]{natbib}
\usepackage{url} 
\usepackage{caption}
\usepackage{subcaption}
\usepackage{ifpdf}
\usepackage{listings}
\usepackage{setspace}
\usepackage{stmaryrd}
\RequirePackage[OT1]{fontenc}
\usepackage{algorithm}
\usepackage{algorithmic}
\usepackage[margin=1in]{geometry}
\usepackage{paralist}

\usepackage{amsbsy}

\usepackage[mathscr]{eucal}

\usepackage{bm}

\usepackage{xspace}

\mathchardef\mhyphen="2D

\ifpdf
\usepackage[colorlinks,urlcolor=blue,citecolor=blue,linkcolor=blue]{hyperref}
\else
\newcommand{\href}[2]{{#2}}
\usepackage{url}
\fi

\ifpdf
\newcommand{\Sec}[1]{\hyperref[sec:#1]{Section~\ref*{sec:#1}}} 
\newcommand{\App}[1]{\hyperref[sec:#1]{Appendix~\ref*{sec:#1}}} 
\newcommand{\Supp}[1]{\hyperref[sec:#1]{Supplement~\ref*{sec:#1}}} 
\newcommand{\Eqn}[1]{\hyperref[eq:#1]{{\rm (\ref*{eq:#1})}}} 
\newcommand{\Part}[1]{\hyperref[part:#1]{(\ref*{part:#1})}} 
\newcommand{\Fig}[1]{\hyperref[fig:#1]{Figure~\ref*{fig:#1}}} 
\newcommand{\Tab}[1]{\hyperref[tab:#1]{Table~\ref*{tab:#1}}} 
\newcommand{\Thm}[1]{\hyperref[thm:#1]{Theorem~\ref*{thm:#1}}} 
\newcommand{\Lem}[1]{\hyperref[lem:#1]{Lemma~\ref*{lem:#1}}} 
\newcommand{\Prop}[1]{\hyperref[prop:#1]{Proposition~\ref*{prop:#1}}} 
\newcommand{\Cor}[1]{\hyperref[cor:#1]{Corollary~\ref*{cor:#1}}} 
\newcommand{\Def}[1]{\hyperref[def:#1]{Definition~\ref*{def:#1}}} 
\newcommand{\Alg}[1]{\hyperref[alg:#1]{Algorithm~\ref*{alg:#1}}} 
\newcommand{\Ex}[1]{\hyperref[ex:#1]{Example~\ref*{ex:#1}}} 
\newcommand{\As}[1]{\hyperref[as:#1]{Assumption~{\rm\ref*{as:#1}}}} 
\newcommand{\Reg}[1]{\hyperref[as:#1]{Condition~\ref*{reg:#1}}} 
\newcommand{\AlgLine}[2]{\hyperref[alg:#1]{line~\ref*{line:#2} of Algorithm~\ref*{alg:#1}}}
\newcommand{\AlgLines}[3]{\hyperref[alg:#1]{lines~\ref*{line:#2}--\ref*{line:#3} of Algorithm~\ref*{alg:#1}}}
\else
\newcommand{\Sec}[1]{{Section~\ref{sec:#1}}} 
\newcommand{\App}[1]{{Appendix~\ref{sec:#1}}} 
\newcommand{\Supp}[1]{{Supplement~\ref{sec:#1}}} 
\newcommand{\Eqn}[1]{{(\ref{eq:#1})}} 
\newcommand{\Part}[1]{{(\ref{part:#1})}} 
\newcommand{\Fig}[1]{{Figure~\ref{fig:#1}}} 
\newcommand{\Tab}[1]{{Table~\ref{tab:#1}}} 
\newcommand{\Thm}[1]{{Theorem~\ref{thm:#1}}} 
\newcommand{\Lem}[1]{{Lemma~\ref{lem:#1}}} 
\newcommand{\Prop}[1]{{Proposition~\ref{prop:#1}}} 
\newcommand{\Cor}[1]{{Corollary~\ref{cor:#1}}} 
\newcommand{\Def}[1]{{Definition~\ref{def:#1}}} 
\newcommand{\Alg}[1]{{Algorithm~\ref{alg:#1}}} 
\newcommand{\Ex}[1]{{Example~\ref{ex:#1}}} 
\newcommand{\Reg}[1]{{R~\ref*{reg:#1}}} 
\fi

\newcommand{\std}{\operatorname{std}}

\newcommand{\T}[1]{{\bm{\mathscr{\MakeUppercase{#1}}}}}

\newcommand{\Real}{\mathbb{R}}

\newcommand{\Ltwo}{\operatorname{L_2}}
\newcommand{\LtwoE}{\operatorname{L_2E}}

\newcommand{\Tra}{^{\sf T}} 
\def\vec{\mathop{\rm vec}\nolimits}

\newcommand{\diag}{\mathrm{diag}}

\newcommand{\V}[1]{{\bm{\mathbf{\MakeLowercase{#1}}}}} 
\newcommand{\VE}[2]{\MakeLowercase{#1}_{#2}} 

\newcommand{\M}[1]{{\bm{\mathbf{\MakeUppercase{#1}}}}} 
\newcommand{\ME}[2]{\MakeLowercase{#1}_{#2}} 

\newcommand{\Mtilde}[1]{{\bm{\tilde \mathbf{\MakeUppercase{#1}}}}} 
\newcommand{\Mn}[2]{\M{#1}^{(#2)}} 
\newcommand{\Mz}[2]{\M{#1}_{(#2)}} 

\newcommand{\KT}[1]{\left\llbracket #1 \right\rrbracket} 
\newcommand{\KTsmall}[1]{\llbracket #1 \rrbracket} 

\newcommand{\amp}{\mathop{\:\:\,}\nolimits}

\newtheorem{theorem}{Theorem}[section]

\newcommand{\blind}{1}

\newenvironment{tightcenter}{%
  \setlength\topsep{0pt}
  \setlength\parskip{0pt}
  \begin{center}
}{%
  \end{center}
}
\begin{document}

\def\spacingset#1{\renewcommand{\baselinestretch}%
{#1}\small\normalsize} \spacingset{1.45}

\if1\blind
{
  \title{\bf Robust Low-rank Tensor Decomposition \\with the L$_2$ Criterion}
  \author{\textbf{Qiang Heng\footnote{Department of Statistics, North Carolina State University},\hspace{2mm}  Eric C.\@ Chi\footnote{Department of Statistics, Rice University},\hspace{2mm} and Yufeng Liu\footnote{Department of Statistics and Operations Research, The University of North Carolina at Chapel Hill}}}
    \date{}
  \maketitle
} \fi

\if0\blind
{
  \bigskip
  \bigskip
  \bigskip
  \begin{center}
    {\LARGE\bf Bayesian Trend Filtering via \\Epigraph-based Proximal MCMC}
\end{center}
  \medskip
} \fi
\begin{abstract}\noindent
    The growing prevalence of tensor data, or multiway arrays, in science and engineering applications motivates the need for tensor decompositions that are robust against outliers. In this paper, we present a robust Tucker decomposition estimator based on the L$_2$ criterion, called the Tucker-$\LtwoE$. Our numerical experiments demonstrate that Tucker-$\LtwoE$ has empirically stronger recovery performance in more challenging high-rank scenarios compared with existing alternatives. The appropriate Tucker-rank can be selected in a data-driven manner with cross-validation or hold-out validation. The practical effectiveness of Tucker-$\LtwoE$ is validated on real data applications in fMRI tensor denoising, PARAFAC analysis of fluorescence data, and feature extraction for classification of corrupted images.
\end{abstract}

\noindent%
\begin{tightcenter}
\end{tightcenter}
\section{Introduction}
There has been growing interest in tensors, or multi-way arrays, since many real-world datasets have a multi-dimensional structure that is not well exploited by two-dimensional matrix-based data analysis. Some of the most important tensor-based data analysis tools are low-rank tensor decompositions, which primarily take two forms:  the CANDECOMP/PARAFAC (CP) decomposition \citep{carroll1970analysis,harshman1970foundations} and the Tucker decomposition \citep{tucker1966some}. In the last decade, several new tensor decomposition paradigms based on alternative notions of tensor rank have been proposed, including low-tubal-rank factorization \citep{kilmer2011factorization}, tensor-train decomposition \citep{oseledets2011tensor} and tensor ring decomposition \citep{zhao2016tensor}. Each decomposition has suitable applications as well as limitations. 

A major challenge of low-rank tensor decomposition is that the observed tensor may be grossly corrupted with outliers or sparse noise. This paper addresses the robust tensor decomposition problem when the underlying tensor has low Tucker-rank. When an $N$-way data tensor $\T{X} \in \Real^{I_1 \times I_2 \times \cdots \times I_N}$ is fully observed, we assume that $\T{X}$ is generated from the following model
\begin{eqnarray*}
\T{X} & = & \T{L} + \T{S} + \T{E},
\end{eqnarray*}
where $\T{L}$ denotes an underlying low Tucker-rank tensor, $\T{S}$ denotes a sparse tensor of outlying entries, and $\T{E}$ denotes a tensor of dense noise.

We also consider the case when $\T{X}$ is only partially observed over a subset $\Omega$ of its indices. Let $[I]$ denote the set of consecutive integers $\{1, \ldots, I\}$. Then the set $\Omega \subset [I_1] \times \cdots \times [I_N]$ is an index set of observed entries, and we assume that 
\begin{eqnarray*}
\ME{X}{i_1i_2\cdots i_N} & = & \ME{L}{i_1i_2\cdots i_N} + \ME{S}{i_1i_2\cdots i_N} + \ME{E}{i_1i_2\cdots i_N}, 
\end{eqnarray*}
for $(i_1, i_2, \cdots, i_N) \in \Omega$. 

Our goal is to recover the latent factors of the underlying low-rank tensor $\T{L}$. Ideally, a robust method should remain effective in the absence of $\T{S}$ or $\T{E}$. If the goal is estimating the low-rank tensor $\T{L}$ instead of its latent factors, we refer to robust tensor decomposition as robust tensor recovery. We focus on the Tucker decomposition in this paper. Since the CP decomposition is a special case of the Tucker decomposition when the CP-rank does not exceed any of the tensor dimensions, the proposed method may also be applied to denoise or reconstruct low CP-rank tensors.

Many CP and Tucker decomposition methods have been proposed in the literature. We discuss these in \Sec{rw}. Our robust formulation adapts the $\LtwoE$ method \citep{scott2001parametric,scott2009l2e} to the  Tucker decomposition. 
The $\LtwoE$ is a minimum distance estimator that minimizes the integrated squared error (ISE) for parametric estimation. The integrated squared error is also referred to as the $\Ltwo$ criterion \citep{hjort1994minimum,terrell1990linear} in nonparametric density estimation, hence the name $\LtwoE$. Consequently, we call our formulation of Tucker decomposition the Tucker-$\LtwoE$. Minimum distance estimators are well known to possess robustness properties \citep{donoho1988automatic}. Moreover, minimization of the $\Ltwo$ criterion has been employed in developing a wide range of robust statistical models including structured sparse models \citep{lozano2016minimum, Chi2022, Liu2023}, quantile regression \citep{Lane2012}, mixture models \citep{Lee2010}, classification \citep{ChiScott2014}, forecast aggregation \citep{Ramos2014}, and survival analysis \citep{Yang2013}.  It also has successes in engineering applications including signal processing tasks such as wavelet-based image denoising \citep{Scott2006} and image registration \citep{ma2013robust, ma2015robust, yang2017remote}. The $\LtwoE$ is attractive among minimum distance estimators since it strikes a good balance between robustness, efficiency, and computational tractability  \citep{scott2001parametric,scott2009l2e}.

The rest of this article is organized as follows. In \Sec{tp}, we introduce tensor notation and terminology. In \Sec{rw}, we briefly review several prominent CP and Tucker decomposition formulations, both non-robust and robust, in the existing literature. 
In \Sec{method}, we present our formulation of robust Tucker decomposition and its solution algorithm. In \Sec{ne} and \Sec{rda}, we demonstrate the practical effectiveness of our approach and its advantage over existing methods in terms of recovery capability with numerical experiments and real data applications. 

\section{Background on Tensors and their Decompositions}
\label{sec:tp}
We review basic operations on matrices and tensors using the terminology and notation in \citet{kolda2009tensor}. 
A tensor $\T{X}\in \Real^{I_1\times I_2\times\dots\times I_N}$ is an element of the tensor product of $N$ real vector spaces. The number of dimensions, or ways, $N$ is called the \textit{order} of tensor $\T{X}$.  Each dimension is also called a \textit{mode}. A \textit{fiber} of $\T{X}$ is a column vector subset of $\T{X}$, defined by fixing all but one of the indices. For a matrix, an order-2 tensor, a mode-1 fiber is a matrix column, and a mode-2 fiber is a matrix row. A \textit{slice} of $\T{X}$ is a matrix subset of $\T{X}$, defined by fixing all but two of the indices.



\subsection{Basic Tensor Operations}

It is often convenient to reshape a tensor into a matrix or a vector. The former is referred to as \textit{matricization}, while the latter is referred to as \textit{vectorization}. The mode-$n$ matricization of a tensor $\T{X} \in \Real^{I_1 \times \cdots \times I_N}$, denoted $\Mz{X}{n} \in \Real^{I_n \times I_{-n}}$ with $I_{-n} = \prod_{k=1, k\neq n}^N I_{k}$, arranges the mode-$n$ fibers as the columns of the matrix $\Mz{X}{n}$ in the following lexicographic order. The tensor element $x_{i_1, \ldots, i_N}$ is mapped to the matrix element of $\Mz{X}{n}$ with index $(i_n, j)$ where $j = 1 + \sum_{k=1, k\not = n}^N (i_k - 1)J_k$ and 
\begin{eqnarray*}
J_k & = & \left\{ \begin{array}{ll} 1 & \text{if } k=1\text{ or if }k=2\text{ and }n=1,\\ \prod_{k'=1, k'\not=i}^{k-1} I_{k'} & \text{otherwise.} \end{array} \right.
\end{eqnarray*}
The vectorization of $\T{X}$, denoted as $\vec(\T{X})$, is the vector obtained by stacking the columns of its mode-1 matricization $\Mz{X}{1}$ on top of each other.

We will use two kinds of products involving tensors and matrices throughout this paper. The elementwise \textit{Hadamard product} of two tensors $\T{X}$ and $\T{Y}$ of identical size $I_1 \times \cdots \times I_N$ is denoted by $\T{X} * \T{Y}$ and is the tensor whose $(i_1, \dots, i_N)$-th element is given by 
$x_{i_1i_2\dots i_N}y_{i_1i_2\dots i_N}$. The \textit{$n$-mode product} of a tensor $\T{X}\in \Real^{I_1\times I_2\times\dots\times I_N}$ with a matrix $\M{A}\in  \Real^{J\times I_n}$ is denoted by $\T{X} \times_n \M{A}$, which is a tensor of size $I_1\times I_2\times \dots\times I_{n-1} \times J \times I_{n+1}\times\dots\times I_{N}$ with elements
\begin{eqnarray*}
\left (\T{X} \times_n \M{A} \right)_{i_1 \cdots i_{n-1} j i_{n+1}\cdots \iota_N} & = & \sum_{i_n = 1}^{I_n} x_{i_1 i_2 \cdots i_N} \ME{A}{j i_n}
\end{eqnarray*}
for $j \in [J]$. Note that the mode-$n$ matricization of the $n$-mode product $\T{X} \times_n \M{A}$ can be expressed as
\begin{eqnarray*}
\left[\T{X} \times_n \M{A}\right]_{(n)} & = & \M{A} \M{X}_{(n)}.
\end{eqnarray*}

The \textit{Frobenius norm} and \textit{$\ell_1$-norm} of a tensor $\T{X} \in \Real^{I_1 \times \cdots \times I_N}$ are  defined as 
\begin{eqnarray*}
\lVert\T{X}\rVert_{\text{F}} & = & \sqrt{\sum_{i_1=1}^{I_1}\sum_{i_2=1}^{I_2}\dots \sum_{i_N=1}^{I_N} x_{i_1i_2\dots i_N}^2} \quad\quad \text{and} \quad\quad
\lVert\T{X}\rVert_1 \amp = \amp  \sum_{i_1=1}^{I_1}\sum_{i_2=1}^{I_2}\dots \sum_{i_N=1}^{I_N} |x_{i_1i_2\dots i_N}|.
\end{eqnarray*}

Finally, we use $\T{X}^{*2}$ to denote the tensor obtained by raising each entry of $\T{X}$ to the power of $2$. We use $a\T{X}+b$ to denote the tensor obtained by multiplying every entry of $\T{X}$ by $a$ and then adding $b$ to every element of the resulting scaled tensor. We denote the sum of all tensor entries as $\operatorname{sum}(\T{X})$ and the elementwise exponential of a tensor as $\exp(\T{X})$.

\subsection{Tensor decompositions and ranks}
The Tucker decomposition of $\T{X}\in \Real^{I_1\times I_2\times\dots\times I_N}$ with rank $R=(r_1,r_2,\dots,r_N)$ aims to find a \textit{core tensor} $\T{G}\in \Real^{r_1\times r_2\times\dots\times r_N}$ and \textit{factor matrices} $\M{A}^{(n)}\in\Real^{I_n\times r_n}$ for $n \in [N]$ such that
\begin{eqnarray*}
\T{X} & \approx & \T{G}\times_1 \M{A}^{(1)} \times_2 \M{A}^{(2)} \times_3 \dots \times_N \M{A}^{(N)} \amp = \amp \KTsmall{\T{G}; \Mn{A}{1}, \dots, \Mn{A}{N}},
\end{eqnarray*}
where the equality uses the more compact notation $\KTsmall{\T{G}; \Mn{A}{1}, \dots, \Mn{A}{N}}$ introduced in  \cite{Kolda2006}. 
Sometimes the columns of $\M{A}^{(n)}$ are required to be orthogonal so that the columns of $\Mn{A}{n}$ can be interpreted as the principal components of the $n$-th mode, but we do not require this in this work. The tensor $\T{X}$ is said to have \textit{Tucker-rank} $R=(r_1,r_2,\dots,r_N)$ if $\operatorname{rank}(\M{X}_{(n)})=r_n$ for $n \in [N]$. 


The CP decomposition for $\T{X}$ with rank $R=r$ aims to find $\V{a}^{(n)}_i \in \Real^{I_n}$ for $n \in [N], i \in [r]$, and a weight vector $\V{\gamma} \in \Real^N$ such that
\begin{eqnarray*}
\T{X} & \approx & \sum_{i=1}^r \gamma_i \V{a}^{(1)}_i \circ \V{a}^{(2)}_i \circ \dots\circ \V{a}^{(N)}_i,
\end{eqnarray*}
where $\circ$ denotes the outer product. Just as the outer product of two vectors yields a rank-1 matrix, the outer product of $N$ vectors yields an $N$-way rank-1 tensor. Thus, the CP model aims to approximate a tensor with a linear combination of rank-1 tensors. Following \cite{kolda2009tensor}, we write the linear combination of rank-1 tensors $\sum_{i=1}^r \gamma_i \V{a}^{(1)}_i \circ \V{a}^{(2)}_i \circ \dots \circ \V{a}^{(N)}_i$ more compactly as $\KTsmall{\V{\gamma}; \Mn{A}{1}, \dots, \Mn{A}{N}}$, 
where $\M{A}^{(n)} = \begin{bmatrix}\V{a}^{(n)}_1\;\V{a}^{(n)}_2\;\dots\; \V{a}^{(n)}_r\end{bmatrix}\in \Real^{I_n\times r}$ are the \textit{CP factor matrices}. The tensor $\T{X}$ is said to have \textit{CP-rank} $r$ if  $r$ is the smallest integer possible for the approximation to hold with equality.  When $r\le \min\{I_1,I_2,\cdots,I_N\}$, the CP decomposition can be viewed as a special case of Tucker decomposition. This is because if $\T{G}$ has dimension $(r,r,\dots,r)$ and is ``superdiagonal," i.e., its only nonzero entries are $g_{ii\dots i}$ for $i \in [r]$, then
\begin{eqnarray*}
\T{G}\times_1 \M{A}^{(1)} \times_2 \M{A}^{(2)} \times_3 \dots \times_N \M{A}^{(N)} & = & \KTsmall{\V{g}; \Mn{A}{1}, \Mn{A}{2}, \dots, \Mn{A}{N}},
\end{eqnarray*}
where $\V{g}\in \Real^{r}$ is a vector containing the  superdiagonal of nonzero entries of $\T{G}$.  A tensor with CP-rank $r$ has Tucker-rank $(r,r,\dots,r)$, but the converse does not hold in general.

For notational simplicity, we will often ``absorb" the weight vector into one of the factor matrices when writing the CP model, e.g.,
\begin{eqnarray*}
\KT{\V{\gamma}; \Mn{A}{1}, \M{A}^{(2)}, \dots, \Mn{A}{N}} & = &
\KT{\Mtilde{A}^{(1)}, \M{A}^{(2)}, \dots, \M{A}^{(N)}},
\end{eqnarray*}
where $\Mtilde{A}^{(1)} = \Mn{A}{1}\diag(\V{\gamma})$ and $\diag(\V{\gamma})$ is the diagonal matrix with $i$-th diagonal entry $\VE{\gamma}{i}$.



\section{Related Work}\label{sec:rw}
\paragraph{Tensor decompositions based on least squares} Non-robust CP and Tucker decompositions are formulated as the solutions to nonlinear least squares problems that minimize the Frobenius norm of the residual tensor. Formally, the CP decomposition solves the following optimization problem
\begin{eqnarray}\label{eq:cp_ls}
\underset{\Mn{A}{1}, \ldots, \Mn{A}{n}}{\operatorname{minimize}} & \left\lVert \T{X} - 
\KT{\Mn{A}{1}, \Mn{A}{2}, \cdots, \Mn{A}{N}}    
\right\rVert_{\text{F}}^2.
\end{eqnarray}
Historically, the alternating least squares (ALS) algorithm \citep{carroll1970analysis,harshman1970foundations} has been the ``work-horse" of solving the above CP decomposition problem, which updates one of the factor matrices while holding the others fixed. \cite{acar2011scalable1}, however, showed that a direct optimization approach can obtain more accurate estimates of the low-rank tensor, especially when the specified rank is greater than the true rank. By direct optimization, we mean that the gradients with respect to the factor matrices are computed and all the factor matrices are updated ``all-at-once" or simultaneously with a local optimization method like nonlinear conjugate gradient method (NCG) or Limited-memory Broyden–Fletcher–Goldfarb–Shanno algorithm (L-BFGS). We refer to this direct optimization approach as CP-OPT. 

Similarly, the Tucker decomposition, or the best rank-$(r_1,r_2,\dots,r_N)$ approximation of $\T{X}$, is formulated as
\begin{eqnarray}\label{eq:tucker_ls}
\underset{\T{G}, \Mn{A}{1}, \dots, \Mn{A}{N}}{\operatorname{minimize}} & \left\lVert \T{X} - \KT{\T{G}; \Mn{A}{1}, \dots, \Mn{A}{N}} \right\rVert_{\text{F}}^2.
\end{eqnarray}
The first method to compute the Tucker decomposition introduced in \cite{tucker1966some} was later shown by \cite{de2000multilinear} to be a generalization of the matrix singular value decomposition (SVD), known today as \textit{Higher-order Singular Value Decomposition} (HOSVD).
However, it does not produce the best fit in terms of relative error. An alternating least squares algorithm named \textit{Higher-order Orthogonal Iteration} (HOOI) \citep{kroonenberg1980principal,kapteyn1986approach,de2000best} has stronger empirical performance and is the most widely adopted method to compute the Tucker decomposition. 
In fact, it has also been shown that HOSVD achieves a sub-optimal rate of estimation error, while HOOI is information-theoretic optimal \citep{zhang2018tensor}.

The above least squares formulations reflect a Gaussian  assumption. This assumption is a natural starting point to develop a robust tensor decomposition formulation. Consequently, our Tucker-$\LtwoE$ method is derived under it. Nonetheless, it is important to note that there have been recent works extending tensor decomposition under non-Gaussian modeling assumptions. For example, \cite{hong2020generalized} studied generalized CP decomposition with various statistically motivated loss functions. \cite{han2022optimal} studied generalized low Tucker-rank tensor estimation, which establishes an upper bound for statistical error and a linear computational convergence rate.

\paragraph{Robust Principal Component Analysis (RPCA)} Perhaps the most classic robust matrix recovery method is \textit{Principal Component Pursuit} \citep{candes2011robust}, which decomposes a corrupted matrix $\M{X}\in\Real^{m\times n}$ as the sum of a low-rank matrix $\M{L}$ and a matrix of sparse outliers $\M{S}$. This is achieved by solving the following convex optimization problem. 
\begin{align}
\begin{aligned}\label{eq:rpca}
\underset{\M{X},\M{S}}{\operatorname{minimize}} \quad & \lVert \M{L}\rVert_*+\lambda\lVert \M{S}\rVert_1 \quad\quad \text{subject to}\quad \M{L} + \M{S} = \M{X},
\end{aligned}
\end{align}
where $\lVert \cdot \rVert_*$ denotes the nuclear norm and $\lambda$ is a nonnegative tuning parameter. 
\cite{candes2011robust} proved that \Eqn{rpca} achieves exact recovery of the low-rank component $\M{L}$ under low-rank and incoherence assumptions. Since the matricizations of a low Tucker-rank tensor are low-rank matrices, RPCA is often used as a baseline for robust tensor recovery. 
\paragraph{Higher-order RPCA} \cite{goldfarb2014robust} proposed \textit{Higher-order Robust Principal Component Analysis} (HoRPCA) as a generalization of RPCA to tensors. HoRPCA comes in several 
different variations. We discuss the three best-performing
variants in this section. The first variant is the \textit{singleton model} (HoRPCA-S), formulated as 
\begin{align}
\begin{aligned}\label{eq:horpcas}
\underset{\T{L},\T{S}}{\operatorname{minimize}}\quad& \sum_{i=1}^N \lVert \M{L}_{(i)}\rVert_*+\lambda\lVert \T{S}\rVert_1
\quad\quad \text{subject to}\quad \T{L}+\T{S} = \T{X}.
\end{aligned}
\end{align}
HoRPCA-S minimizes the sum of nuclear norms of all the  matricizations of $\T{L}$ to encourage each mode to be low-rank. The descriptor ``singleton" is in contrast with the \textit{mixture model} (HoRPCA-W), formulated as 
\begin{align}
\begin{aligned}\label{eq:horpcaw}
\underset{\T{L}_i,\T{S}}{\operatorname{minimize}}\quad& \sum_{i=1}^N \lVert \M{L}_{i,(i)}\rVert_*+\lambda\lVert \T{S}\rVert_1
\quad\quad \text{subject to}\quad \sum_{i=1}^N\T{L}_i+\T{S} = \T{X}.
\end{aligned}
\end{align}
HoRPCA-W represents the underlying tensor as the sum of $N$ tensors that are only low-rank in one mode. \cite{tomioka2011statistical} first introduced the mixture model which can be considered a relaxation of the singleton model. \cite{yang2015robust} later  proposed robust tensor recovery also using the mixture model along with more robust loss functions. The mixture model can automatically detect the rank-deficient modes and yields better recovery results when the underlying tensor is only low-rank in certain modes. However, in our experience, the limitation of the mixture model is that it does not approximate the low-rank tensor well when the minimum rank in the Tucker rank tuple is relatively large. The variant with the strongest recovery performance presented in \cite{goldfarb2014robust} is the \textit{constrained nonconvex model} (HoRPCA-C), formulated as 
\begin{align}
\begin{aligned}\label{eq:horpcac}
\underset{\T{L},\T{S}}{\operatorname{minimize}}\quad& \lVert \T{S}\rVert_1
\quad\quad \text{subject to}\quad \T{L}+\T{S} = \T{X},\quad \operatorname{rank}(\M{L}_{(i)})\le r_i,\; i \in [N].
\end{aligned}
\end{align}
The key difference between the formulation in \Eqn{horpcac} and those in \Eqn{horpcas} and \Eqn{horpcaw} is that \Eqn{horpcac} enforces a hard constraint on the Tucker-rank $(r_1,r_2,\dots,r_N)$ whereas the other formulations trade off the rank of the latent tensor $\T{L}$ with the $\ell_1$-norm of the outlier tensor $\T{S}$ via the penalty parameter $\lambda$. 

\cite{goldfarb2014robust} iteratively solves \Eqn{horpcas}, \Eqn{horpcaw} and \Eqn{horpcac} using 
the alternating direction method of multipliers (ADMM) algorithm \citep{boyd2011distributed}. In particular, since \Eqn{horpcac} is nonconvex, the standard convergence guarantees of ADMM for convex programs do not apply. However, \Eqn{horpcac} demonstrates strong empirical convergence in practice.  
\paragraph{Bayesian Robust Tensor Factorization} \cite{zhao2015bayesian} approached the robust CP decomposition problem in a generative manner with \textit{Bayesian Robust Tensor Factorization.} (BRTF). Normal-gamma priors are used to induce column sparsity of the factor matrices and elementwise sparsity of the outliers. We direct readers to \cite{zhao2015bayesian} for the detailed hierarchical model setup. What makes BRTF attractive is that it can  automatically infer the appropriate CP-rank.

\paragraph{Riemannian Gradient Descent} Recently, \cite{cai2022generalized} introduced a general framework under a  low-rank plus sparse tensor model. The algorithm is based on Riemannian gradient descent and a novel gradient pruning procedure, which is able to estimate both the low-rank tensor and the outlying sparse tensor. The appropriate Tucker-rank and sparsity level of outliers can be tuned with a BIC-type criterion. Performance bounds for both the low-rank and the sparse tensors are established under suitable conditions. The proposed algorithm is also applicable to Bernoulli and Poisson distributed data. We refer to the algorithm described in  \cite{cai2022generalized} as RGrad in this article.

\paragraph{Partial observations} Real-world tensor data is often not fully observed. In the tensor completion literature, a binary weight tensor $\T{W}$ whose entries are 0/1 to indicate missing or observed is often used as a mask to model missing entries. For example, CP and Tucker decomposition in the presence of missing data can be formulated as
\begin{eqnarray}\label{eq:cp_wopt}
\underset{\Mn{A}{1}, \dots, \Mn{A}{N}}{\operatorname{minimize}} & \left\lVert \T{W}*\left(\T{X} - \KT{\Mn{A}{1}, \dots, \Mn{A}{N}} \right)\right\rVert_{\text{F}}^2,
\end{eqnarray}
and
\begin{eqnarray}\label{eq:tucker_wopt}
\underset{\T{G},\Mn{A}{1}, \dots, \Mn{A}{N}}{\operatorname{minimize}} & \left\lVert\T{W}*\left( \T{X} - \KT{\T{G}; \Mn{A}{1}, \dots, \Mn{A}{N}} \right )
 \right\rVert_{\text{F}}^2.
\end{eqnarray}
Similar to \Eqn{cp_ls} and \Eqn{tucker_ls}, \Eqn{cp_wopt} and \Eqn{tucker_wopt} can also be solved with direct optimization, e.g., CP-WOPT \citep{acar2011scalable2} and Tucker-WOPT \citep{filipovic2015tucker}. For RPCA and HoRPCA, an easy way to deal with missing data is to enforce the equality constraints only on observed entries. Similarly for BRTF,  we can choose to incorporate only the observed tensor entries into the hierarchical model.

It is worth mentioning that there is significant interest in tensor completion in recent research. In the last few years there are many advances on the theoretical front for low CP/Tucker-rank tensor completion. For example, \cite{cai2019nonconvex} studied the reconstruction
of a low-rank symmetric tensor and proposed a
two-stage nonconvex algorithm which achieves optimal $\ell_\infty$ statistical accuracy. Building upon \cite{cai2019nonconvex}, \cite{cai2020uncertainty} studied the nonconvex tensor completion problem from an uncertainty quantification perspective. \cite{xia2019polynomial} studied the sample size requirement for the exact recovery of a low-rank tensor from a subset of its entries, using a spectral initialization method and gradient descent. \cite{zhang2019cross} proposed a novel tensor measurement scheme for low-rank tensor completion. \cite{xia2021statistically} proposed a procedure for low-rank tensor completion from noisy entries based on spectral initialization and power iteration that is computationally efficient and achieves the optimal rates of convergence. Recently, \cite{tong2022scaling} developed a scaled gradient descent approach to low-rank tensor completion and regression which converges at a linear rate independent of the condition number of the true tensor.

\paragraph{Remark} Among the previously reviewed methods, RPCA, HoRPCA-S and HoRPCA-W induce a low-rank structure using nuclear norm penalties while CP-OPT, HOOI, HoRPCA-C and RGrad require the rank to be explicitly specified. We categorize methods that employ nuclear norm penalties as ``penalized formulations" and methods requiring specification of rank as ``rank-constrained formulations." Our approach, Tucker-$\LtwoE$, is an instance of the latter. In \Sec{ne}, we demonstrate that rank-constrained formulations are generally more robust and handle dense noise better than penalized formulations, provided that the ranks are appropriately specified. We also illustrate in \Sec{rm} that HoRPCA-C and Tucker-$\LtwoE$ can tolerate some level of rank overestimation if the noise is sparse.

We acknowledge several other robust CP/Tucker decomposition methods that we have not detailed in this section due to the limitation of space. \cite{anandkumar2016tensor}, for example, proposed an iterative thresholding algorithm for robust tensor decompositions which is designed to recover CP models with orthogonal factors. \cite{gu2014robust} studies the statistical performance of a convex formulation of robust tensor decomposition. \cite{wu2017robust} uses the Cauchy distribution to handle long-tail noise in CP and Tucker decomposition.

\section{Methodology}\label{sec:method}
In this section, we introduce our proposed Tucker-$\LtwoE$ method.  
We briefly review the $\LtwoE$ method in \Sec{l2em} and then develop Tucker-$\LtwoE$ in \Sec{tuckerl2em}. The 
algorithm and implementation details for Tucker-$\LtwoE$ can be found in \Sec{algorithm}.
\subsection{The $\LtwoE$ Method}\label{sec:l2em}
We first review the parametric estimation framework using the $\Ltwo$ criterion proposed by \cite{scott2001parametric,scott2009l2e}. Let $\phi(x)$ be the unknown true density we aim to estimate and $\tilde{\phi}(x\mid\theta)$ be the density of a member of the family of parametric models specified by the parameter $\theta \in \Theta$. We seek the parameter $\theta$ that minimizes the ISE  between $\phi(x)$ and $\tilde{\phi}(x \mid\theta)$
\begin{eqnarray}
\label{eq:ISE}
\int \left [\tilde{\phi}(x \mid \theta) - \phi(x) \right ]^2 dx.
\end{eqnarray}
Of course, recovering $\theta$ in this way is impossible in practice since $\phi$ is unknown.  Fortunately, although we cannot minimize the L$_2$ distance between $\phi(x)$ and $\tilde{\phi}(x \mid \theta)$ directly, we can minimize an unbiased estimate of the distance.  To do this, we first expand \Eqn{ISE} as
\begin{eqnarray*}
\label{eq:discreteL2E}
\int \tilde{\phi}(x \mid \theta)^2 \,dx - 2\int \tilde{\phi}(x \mid \theta)\,\phi(x)\,dx +  \int \phi(x)^2 \, dx.
\end{eqnarray*}
The second integral is the expectation $\mathbb{E}_{X}[\tilde{\phi}(X \mid \theta)]$, where $X$ is a random variable drawn from $\phi$. Therefore, the sample mean provides an unbiased estimate of this quantity. The third integral does not involve the parameter of interest $\theta$ and may be ignored in the computation of a minimizer. 
The first integral has a closed-form expression for many parametric models. In this work, we assume that $\tilde{\phi}(x\mid\theta)$ is a normal density where $\theta$ consists of a mean and precision (inverse standard deviation) parameter. Under this assumption, the integral $\int \tilde{\phi}(x\mid\theta)^2dx$ can be written as an explicit function of the precision parameter alone. 
As a concrete example, consider the univariate case and assume that $\theta = (\mu, \tau)$ and $\tilde{\phi}(x\mid\theta)$ is the density function of a normal random variable $X \sim \mathcal{N}(\mu,\tau^{-2})$. Then the $\LtwoE$ for the univariate mean and precision is
\begin{eqnarray}\label{eq:l2enormal}
\hat{\theta}_{\LtwoE} &=& \underset{\mu,\tau}{\arg\min}\; h(\mu, \tau),
\end{eqnarray}
where
\begin{eqnarray}\label{eq:unil2e}
h(\mu, \tau) & = &
\frac{\tau}{2\sqrt{\pi}}- \frac{\tau}{n}\sqrt{\frac{2}{\pi}}\sum_{i=1}^n \exp\left(-\frac{\tau^2}{2}(x_i-\mu)^2\right).
\end{eqnarray}

For a fixed $\tau$, the $\mu$ that minimizes \Eqn{unil2e} approaches the MLE of $\mu$ as $\tau$ approaches zero. To see this, note that the Taylor expansion of $-\exp(-t)$ around 0 is $-1+t+o(t)$. Therefore for sufficiently small $\tau$,
\begin{eqnarray*}
h(\mu, \tau) & \approx & \frac{\tau}{2\sqrt{\pi}} - \sqrt{\frac{2}{\pi}}\tau + \frac{\tau^3}{n\sqrt{2\pi}}\sum_{i=1}^n (x_i-\mu)^2. 
\end{eqnarray*}
\begin{figure}[tbp]
  \centering
  \includegraphics[width=\textwidth]{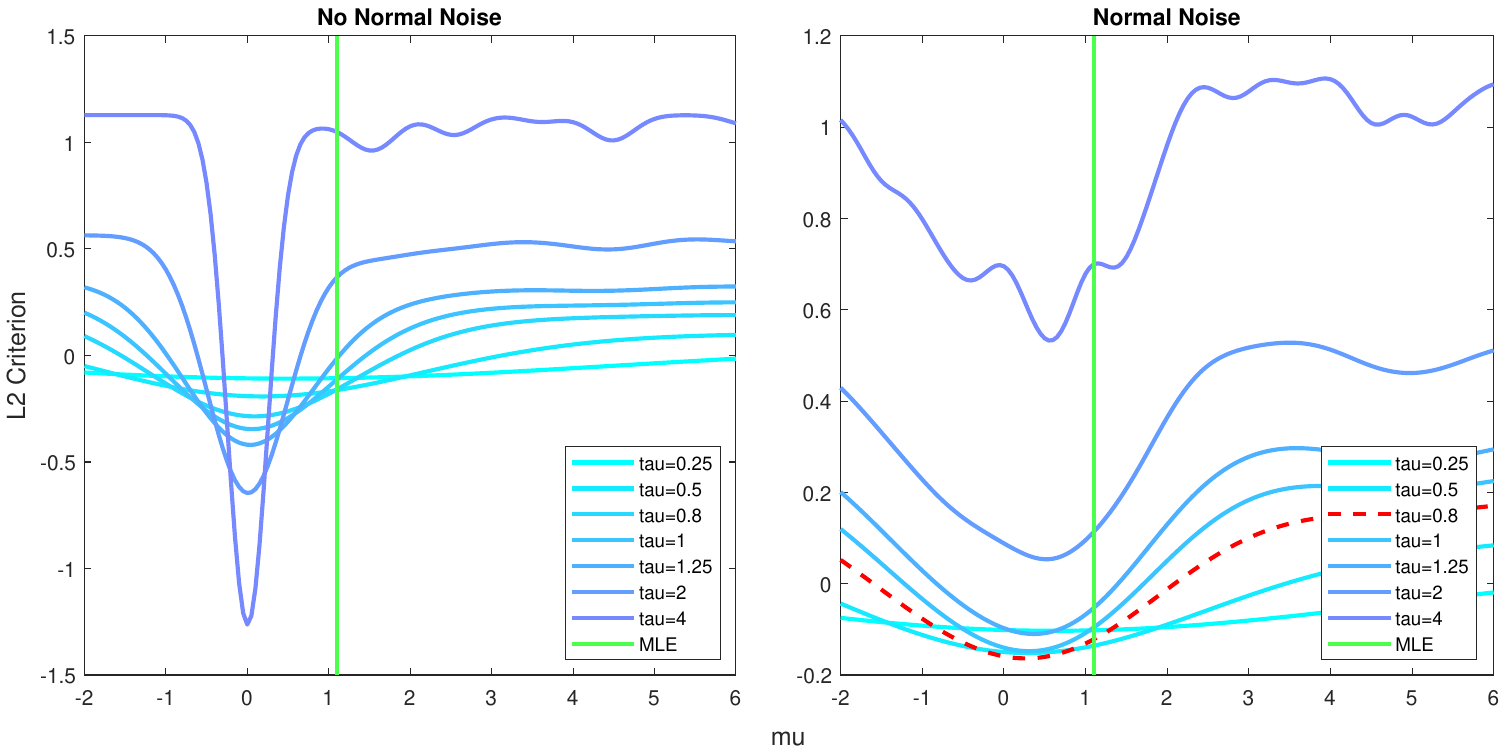}
  \caption{The $\Ltwo$ criterion as a function of $\mu$ with different values of $\tau$. The green vertical line indicates the maximum likelihood estimator of $\mu$, in this case simply $\bar{x}$.}
  \label{fig:l2e}
\end{figure} 
We can visualize how the $\Ltwo$ criterion $h(\mu, \tau)$ varies with $\mu$ for fixed values of $\tau$ to illustrate how the $\LtwoE$ achieves robustness. We consider two examples. In the first example, $\phi(x)$ is a three-to-one mixture of the constant 0 and $\operatorname{Unif}[0,10]$. The second example is identical to the first but the observations are further corrupted with additive  $\mathcal{N}(0,1)$ noise. One hundred observations $x_1, x_2, \dots, x_{100}$ are generated for each example. In both examples, the true value for $\mu$ is 0 and the 25\% uniformly distributed observations can be regarded as outliers. Intuitively, the true value for $\tau$ is $+\infty$ for the first example and $1$ for the second example. \Fig{l2e} shows $h(\mu, \tau)$ as a function of $\mu$ for different values of $\tau$ with the first example in the left panel and the second example in the right panel.

The left panel of \Fig{l2e} shows that the $\mu$ that minimizes $h(\mu, \tau)$ is nearly identical to the MLE of $\mu$ when $\tau$ is very small. As $\tau$ increases, the minimizing $\mu$ becomes closer to the true value 0. In other words, the $\Ltwo$ criterion is less influenced by the outliers and consequently, the $\LtwoE$ for $\mu$ is more robust. We also see that if $\tau$ becomes too large, however, many spurious local minima appear in the optimization landscape, which may cause difficulties for gradient-based local optimization algorithms. 

The right panel of \Fig{l2e} shows that the $\Ltwo$ criterion curve with $\tau=0.8$ (highlighted in red dashed line) attains the smallest minimum value. Moreover, the minimum of $h(\mu, \tau)$ with $\tau=0.8$ is the closest (0.25) to the true value 0. This suggests that \Eqn{l2enormal} is able to automatically choose suitable $\tau$ in the presence of normal noise, although it slightly underestimates the precision parameter. \cite{scott2009l2e} also observed this underestimation phenomenon.

\subsection{Tucker-$\LtwoE$}\label{sec:tuckerl2em}
We set up the optimization problem for estimating the Tucker-$\LtwoE$ model in stages. We start with a natural adaptation of \Eqn{l2enormal} to accommodate tensor data which replaces the location parameter $\mu$ with a latent mean tensor $\T{L}$.
A preliminary formulation of robust tensor estimation based on the $\Ltwo$ criterion is to minimize the following objective function
\begin{eqnarray}
\label{eq:tuckerl2enu}
h\left(\T{L}, \tau \right) & = & 
 \frac{\prod_{n=1}^N I_n}{2\sqrt{\pi}}\tau-\sqrt{\frac{2}{\pi}}\tau\operatorname{sum}\left(\exp\left[-\frac{\tau^2}{2}\left(\T{X} - \T{L} \right)^{*2}\right]\right),
\end{eqnarray}
where $\T{X}\in \Real^{I_1\times I_2\times\dots \times I_N}$ is the observed noisy tensor. 


When the data tensor $\T{X}$ has  missing entries and is observed only on the index set $\Omega$, similar to CP-WOPT and Tucker-WOPT, we can sum over only the observed entries in the objective to account for missing data and minimize the natural generalization of the objective function in \Eqn{tuckerl2enu}
\begin{eqnarray}
\label{eq:tuckerl2enu_missing}
h_\Omega\left(\T{L}, \tau \right) & = & 
\frac{\operatorname{sum}\left(\T{W}\right)}{2\sqrt{\pi}}\tau-\sqrt{\frac{2}{\pi}}\tau\operatorname{sum}\left(\T{W}*\exp\left(-\frac{\tau^2}{2}(\T{X}-\T{L})^{*2}\right)\right).
\end{eqnarray}
Recall that the tensor $\T{W} \in \Real^{I_1\times I_2\times\dots \times I_N}$ is binary and depends on $\Omega$ in the following manner. The $(i_1, i_2, \ldots, i_N)$-th entry of $\T{W}$ is 1 if $\ME{X}{i_1i_2\ldots i_N} \in \Omega$ and is 0 otherwise. Note that $h_\Omega(\T{L}, \tau)$ and  $h(\T{L}, \tau)$ coincide when $\T{X}$ is fully observed, i.e., $\Omega = [I_1]\times \cdots \times [I_N].$ Thus, we will work with the more general objective function $h_\Omega$ moving forward.

 To estimate a low Tucker-rank tensor,
we parameterize $\T{L}$ as $\T{L} = \KT{\T{G}; \Mn{A}{1}, \dots, \Mn{A}{N}}$. Notice that this parameterization is equivalent to imposing the constraints $\operatorname{rank}(\M{L}_{(n)})\le r_n, n \in [N]$ on $\T{L}$ (see \cite{zhang2018tensor}). Thus we seek the solution to the following optimization problem over the parameters $\T{G}, \Mn{A}{1}, \dots, \Mn{A}{N}$ and $\tau$.
\begin{eqnarray*}
\text{minimze}\; 
h_\Omega\left(\KT{\T{G}; \Mn{A}{1}, \dots, \Mn{A}{N}}, \tau \right) \quad\quad \text{subject to}\quad\quad \tau > 0.
\end{eqnarray*}

We now turn our attention to details concerning the parameter $\tau$. 
Notice that $\tau$ must be positive since it is a precision parameter and consequently introduces an additional constraint over an open set. An easy way to ensure a positive precision while at the same time eliminating the strict positivity constraint is to reparameterize $\tau$ as $\tau = \exp(\eta)$ and optimize over $\eta \in \Real$. 
Moreover, recall in \Sec{l2em}, we discussed that although the $\LtwoE$ is more robust when $\tau$ is larger, the accompanying spurious local minima may make computing solutions of \Eqn{tuckerl2enu} harder. In other words, the precision parameter $\tau$ trades-off robustness and the ``roughness" of the optimization landscape. We also see from the left panel of \Fig{l2e} that the minimum value of $h(\mu, \tau)$ always decreases as $\tau$ increases. This suggests that \Eqn{tuckerl2enu} may not even have a finite infimum if we allow $\tau$ to diverge to infinity. Therefore, it is reasonable to impose an upper bound on $\tau$, or equivalently on $\eta$. Thus, we seek the solution to the following optimization problem over the parameters $\T{G}, \Mn{A}{1}, \dots, \Mn{A}{N}$ and $\eta$.
\begin{eqnarray}
\label{eq:tuckerl2e}
\text{minimize}\; 
h_\Omega\left(\KT{\T{G}; \Mn{A}{1}, \dots, \Mn{A}{N}}, e^{\eta} \right) \quad\quad \text{subject to}\quad\quad \eta \leq \eta_{\max}.
\end{eqnarray}

\Thm{well} gives some justification for reparameterizing $\tau$ as $\exp(\eta)$ and placing an upper bound on $\eta$ in \Eqn{tuckerl2e}. 
\begin{theorem}\label{thm:well}
Problem \Eqn{tuckerl2e} has a finite infimum and the infimum is negative.
\end{theorem}

We provide a proof of \Thm{well} in the supplement. \Thm{well} is relevant in our problem setup since it has the following implications in our computation process. First, we have seen in the left panel of \Fig{l2e} that if the noise is sparse, the $\LtwoE$ problem may be ill-posed with no finite infimum. \Thm{well} assures us that if we place an upper bound on $\eta$, we can guarantee a finite infimum. Second, when $\tau=0$, it is not hard to see from \Eqn{tuckerl2enu_missing} that the objective value will be 0. If the infimum of \Eqn{tuckerl2e} is guaranteed to be negative, then any point with $\tau=0$ is suboptimal, thus we do not risk excluding a potential minimizer through reparameterizing $\tau$ as $\exp(\eta)$.

Before discussing how to compute Tucker-$\LtwoE$ next, we emphasize a key feature of our approach is that it  simultaneously optimizes or estimates a precision parameter as well as a latent low-rank tensor. As we saw earlier, this precision parameter controls the cutoff for when an entry is large enough to be effectively trimmed from the model fitting. As far as we are aware, Tucker-$\LtwoE$ is unique in its ability to jointly optimize precision and latent low-rank tensor parameters. Other methods that employ a precision parameter treat it as a hyperparameter and consequently employ a separate estimation or setting procedure for the precision parameter. For example, \cite{wu2017robust} computes an estimate of the precision parameter based on the residuals of the least squares estimate of the low-rank tensor, but this limits the recovery capability of the model as the least squares estimates are of poor quality in challenging high-rank scenarios.

\subsection{Solution Algorithm and Implementation Details}\label{sec:algorithm}
The optimization problem in 
\Eqn{tuckerl2e} involves differentiable functions of all the model parameters with a simple box constraint on $\eta$. The high dimensionality of the parameter space renders second-order algorithms impractical. In contrast, the classic quasi-Newton method L-BFGS-B \citep{liu1989limited,byrd1995limited,zhu1997algorithm} is particularly well suited for solving \Eqn{tuckerl2e}. A detailed derivation of the gradient of the objective in \Eqn{tuckerl2e} with respect to $\M{A}^{(n)}$, $\T{G}$, and $\eta$ can be found in the supplement. Since the optimization problem in \Eqn{tuckerl2e} is nonconvex, initialization is critical. We present two initialization strategies in this article. The first is a simple strategy from \cite{filipovic2015tucker}:
\begin{algorithm}[htbp]
\caption{Mean Imputation + HOOI/HOSVD}\label{alg:init1}
\begin{algorithmic}[1]
\REQUIRE $\T{X},\T{W}\in\Real^{I_1\times I_2 \times \dots \times I_N}, (r_1,r_2,\dots,r_N)$.
\STATE Impute missing entries with the mean of the observed entries of $\T{X}$. 
\STATE Compute $\T{G}_0$ and $\M{A}^{(n)}_0$ via HOOI/HOSVD of $\T{X}$ with rank $(r_1,r_2,\dots,r_N)$.
\RETURN ($\T{G}_0$, $\M{A}^{(n)}_0$).
\end{algorithmic}
\end{algorithm}

The second is a popular initialization procedure in the recent tensor completion literature, which we call spectral initialization with diagonal deletion, see  for example \cite{cai2019nonconvex,xia2019polynomial,xia2021statistically,tong2022scaling}. In our experiments,  we find that this initialization procedure offers some improvement over \Alg{init1}  when the underlying tensor has low CP-rank and a large percentage of tensor entries is missing. However, we note that in most cases, \Alg{init1} works similarly or better, especially when the underlying Tucker-rank is relatively high. Therefore our default initialization procedure is \Alg{init1} in our following experiments unless explicitly specified otherwise. We discuss the details of the second initialization strategy in the supplement.

We find the initial value of $\eta$ to have minimal impact on the solution, consequently we initialize $\eta$ using a small value $\log(0.01)$, where $\log$ is the natural log function. We also observe that numerical issues can occur when the tensor entries are somewhat large in magnitude due to the exponent function in the objective of $\Eqn{tuckerl2e}$. Scientific computing languages compute  $\exp(-x)$ via a power series so that the numeric precision degrades when $x$ becomes larger in absolute value. Therefore as a pre-processing step, the observed tensor entries are rescaled to have a mean absolute deviation (MAD) of 0.1. We revert the estimated tensor to the original scale after the decomposition is complete. Another practical concern is the choice of Tucker-rank $(r_1,r_2,\dots,r_N)$ and the upper bound $\eta_{\max}$. In particular, for the upper bound $\eta_{\max}$, we want to select a value such that we have sufficient robustness but the loss landscape is still smooth enough for L-BFGS-B to find a good solution. In \Sec{rm} and \Sec{fmri}, we demonstrate that Tucker-rank $(r_1,r_2,\dots,r_N)$ can be selected in a data-driven manner using cross-validation or hold-out validation if computation is intensive. For $\eta_{\max}$, we find that $\eta_{\max}=\log(50)$ works well for a wide range of problems and we have used it for all of our experiments except the feature extraction application in \Sec{fe} where $\eta_{\max}$ is set at $\log(20)$ for optimal performance. It can also be tuned along with the Tucker-rank using cross-validation or hold-out validation if warranted or desired.

\begin{algorithm}[htbp]
\caption{Tucker-$\LtwoE$}\label{alg:pseudocode}
\begin{algorithmic}[1]
\REQUIRE $\T{X},\T{W}\in\Real^{I_1\times I_2 \times \dots \times I_N}, (r_1,r_2,\dots,r_N),$ and $\eta_{\max}$.
\STATE Calculate the MAD of the observed entries of $\T{X}$, denoted as $s$.
\STATE Rescale $\T{X}$ as $\frac{1}{10s}\T{X}$.
\STATE Compute initial estimates ($\T{G}_0$, $\M{A}^{(n)}_0$) with \Alg{init1} or spectral initialization with diagonal deletion.
\STATE Set $\eta_0 = \log(0.01)$. 
\STATE  Using \Eqn{tuckerl2e} as the objective and ($\T{G}_0$, $\Mn{A}{n}_0$, $\eta_0$) as the initial value, update ($\T{G}$, $\Mn{A}{n}$, $\eta$) with L-BFGS-B until convergence of objective value or the maximum number of iterations is reached. The final iterate is denoted as ($\T{G}_*$, $\Mn{A}{n}_*$, $\eta_*$).
\RETURN ($10s\T{G}_*$, $\Mn{A}{n}_*$, $\eta_*$). 
\end{algorithmic}
\end{algorithm}

\Alg{pseudocode} summarizes our procedure for computing Tucker-$\LtwoE$. We implement our algorithm in the MATLAB R2021a computing environment. We use the Tensor Toolbox for MATLAB version 3.2.1\footnote{https://www.tensortoolbox.org/} \citep{bader2006algorithm,bader2008efficient} for basic tensor classes and operations. We also use the implementation of HOOI and HOSVD in the Tensor Toolbox to compute the initial estimate of $\T{G}$ and $\M{A}^{(n)}$ in \Alg{init1}. We use the implementation of L-BFGS-B by \cite{becker2015bfgs}\footnote{https://github.com/stephenbeckr/L-BFGS-B-C}. We direct readers to \cite{byrd1995limited} for the algorithmic details of L-BFGS-B. We note that since L-BFGS-B is a gradient-based local optimization algorithm, \Alg{pseudocode} is only able to find a locally optimal or critical point of \Eqn{tuckerl2e}. Additionally, since L-BFGS-B is a descent method \citep{byrd1995limited} and \Thm{well} ensures that our objective in  \Eqn{tuckerl2e} is bounded below, the sequence of objective function evaluations over the iterate sequence  is guaranteed to converge with L-BFGS-B updates.

\section{Numerical Experiments}\label{sec:ne}

We consider both the CP model and the Tucker model in our simulation studies. The tensor dimension is set at $(50,50,50)$.
For the CP model, the entries of the factor matrices are independently drawn from $\mathcal{N}(0,1)$. For the Tucker model, we adopt a similar data generation protocol to \cite{cai2022generalized}.  A tensor of size $(50,50,50)$ with random normal entries is first generated. Then the tensor is truncated to have Tucker-rank $(r,r,r)$ with HOSVD. We use relative error, defined as $\textsc{re} = \lVert \hat{\T{L}} - \T{L} \rVert_{\text{F}}/\lVert\T{L}\rVert_{\text{F}},$ as the primary goodness-of-fit metric. After generating the low-rank tensor $\T{L}$, we randomly select a fraction $\delta$ of the tensor entries to be corrupted with outliers drawn from $\operatorname{Unif}[-M,M]$.  We use a relatively large magnitude $M=10\std(\vec(\T{L}))$ in the following experiments. Optionally, a layer of dense Gaussian noise $\T{E}$ can also be added, whose scale is set such that $\lVert \T{E}\rVert_{\text{F}}/\lVert \T{L}\rVert_{\text{F}} = 0.1$.
We use CP-OPT, HOOI, BRTF, HoRPCA-S, HoRPCA-C and RGrad as the baseline methods in this section. We use the implementation of CP-OPT and HOOI in the Tensor Toolbox. The software of HoRPCA-S, HoRPCA-C and BRTF can be found on the authors' websites. The software of RGrad can be found in the supplementary materials of \cite{cai2022generalized}. We provide code and demo examples for our proposed method at \url{https://github.com/qhengncsu/TuckerL2E}.
\begin{figure}[htbp]
  \centering
  \includegraphics[width=\linewidth]{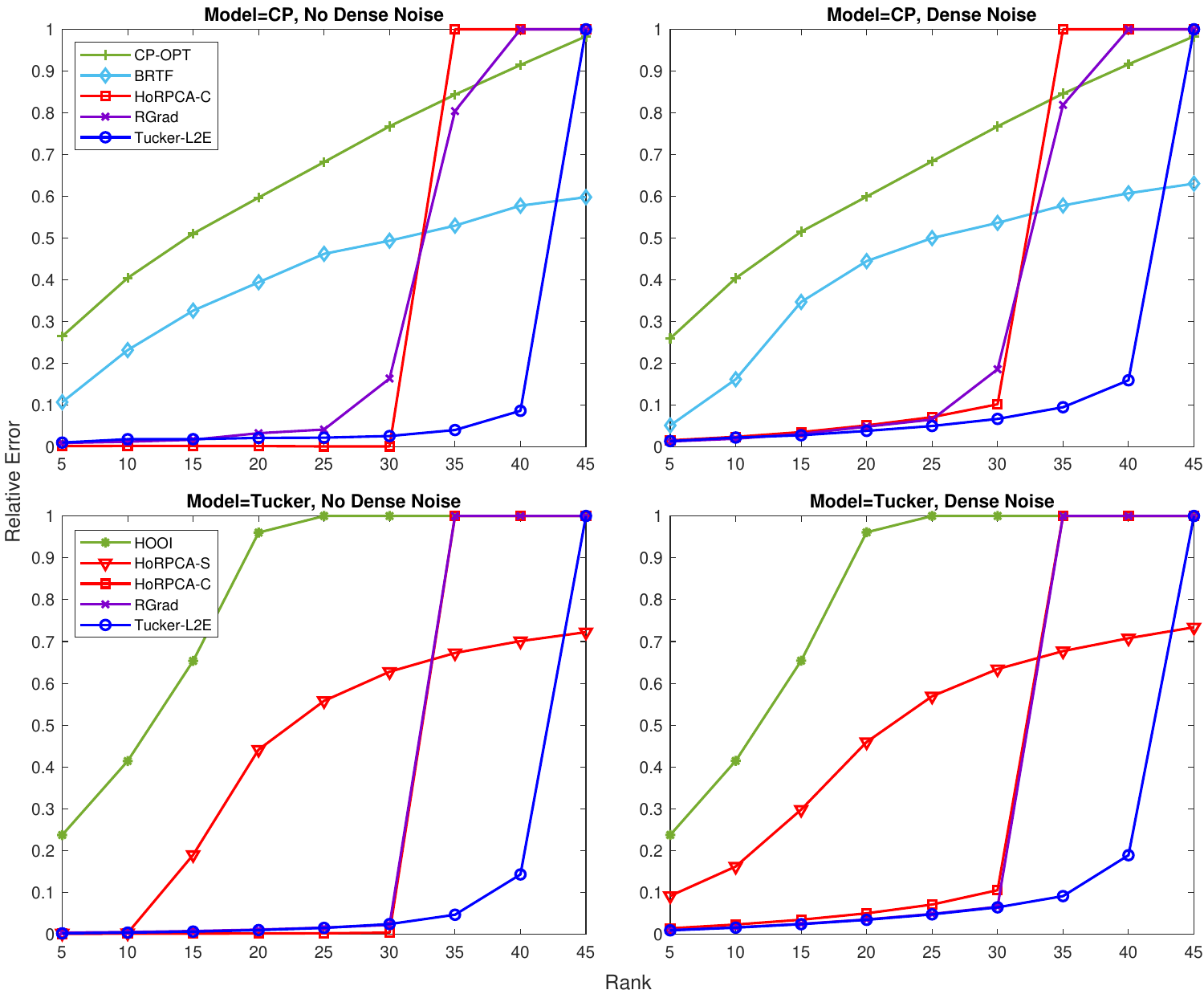}
  \caption{Recovery results on fully observed tensors with increasing CP or Tucker-rank. Outlier sparsity is set at 0.25. Data points are averaged over 50 random replicates. Average relative errors larger than 1 are capped at 1 to suit the display.}
  \label{fig:main}
\end{figure}
\subsection{Evaluating Robustness versus Rank}\label{sec:hr}

The inverse problem of robust Tucker decomposition becomes more challenging as the underlying Tucker rank or outlier percentage increases. In this section, we contrast the recovery performance of the baseline methods and Tucker-$\LtwoE$ by generating third-order tensors with increasing CP or Tucker-rank ($R = 5,10,\dots,45$ or $R=(5,5,5), (10,10,10),\dots, (45,45,45)$), under outlier corruption and in the presence or absence of dense noise. Note that for a tensor with an underlying CP-rank of $r$, we can still compute a Tucker decomposition  with Tucker-rank $(r,r,r)$ to reconstruct the tensor. In this section we keep the outlier sparisty at 25\%. For HoRPCA-S, we tune the penalty parameter with the ground truth. For rank-constrained formulations, the specified CP or Tucker-ranks are set to the true ranks.  We provide additional details of parameter settings in the supplement.

\Fig{main} shows that the convex penalized formulation HoRPCA-S works  well when the tensor rank is low but loses accuracy as the rank increases. HoRPCA-C, RGrad and Tucker-$\LtwoE$ demonstrate competitive recovery performance in most cases, except when the rank is too close to the data dimension, particularly at rank 35-45. Notably, Tucker-$\LtwoE$ appears to be able to tolerate a higher Tucker-rank than HoRPCA-C and RGrad, both in the CP model and the Tucker model. More specifically,  Tucker-$\LtwoE$ is able to provide a reasonably good reconstruction at rank 35 and 40 while HoRPCA-C and RGrad break down.

\subsection{Phase Transition of Rank and Outlier Sparsity}\label{sec:pt}
\begin{figure}[htbp]
  \centering
  \includegraphics[width=0.9\linewidth]{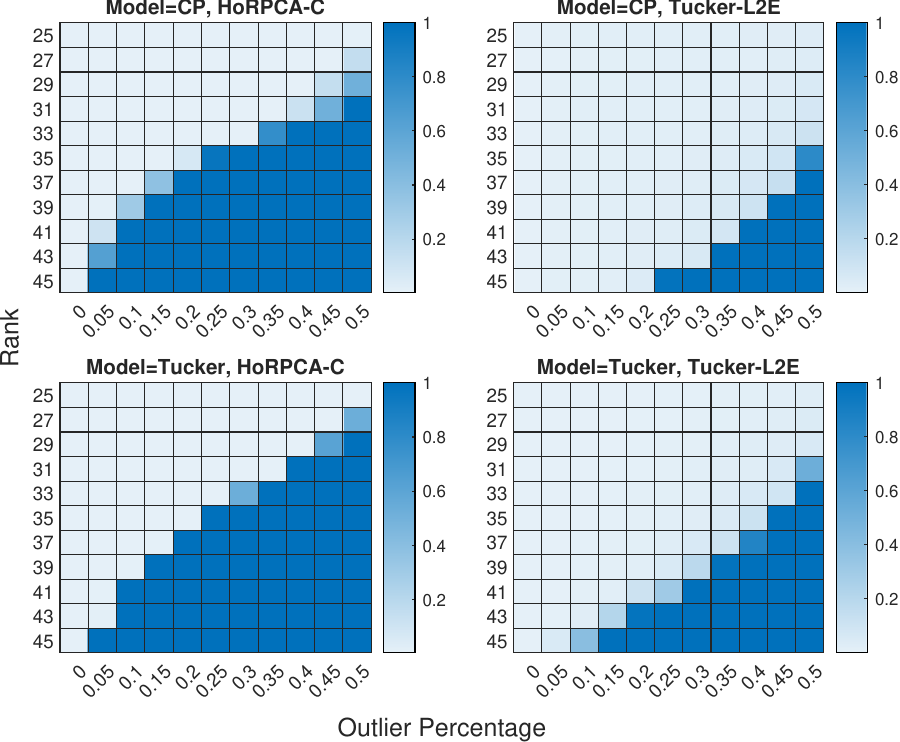}
  \caption{Phase transition diagrams in high-rank scenarios with varying percentages of outliers. Heatmap shows the average relative error of 20 random replicates. Average relative errors larger than 1 are capped at 1 to suit the display.}
  \label{fig:pt}
\end{figure}
In the previous section, we saw that compared with HoRPCA-C and RGrad, Tucker-$\LtwoE$ appears to be able to tolerate a higher rank at the given outlier percentage.  Although RGrad handles dense noise better than HoRPCA-C, HoRPCA-C appears to be a slightly stronger baseline when it comes to stability in high-rank scenarios. To further investigate the phase transition behavior of HoRPCA-C and Tucker-$\LtwoE$, we generate tensors with high rank ($R=25,27,\dots,45$ or $R=(25,25,25),(27,27,27),\dots,(45,45,45)$) and vary the percentage of outliers taking $\delta = 0, 0.05, \dots, 0.5$. Full observations are used and no dense noise is added. The specified Tucker-ranks are set to be equal to the true CP or Tucker-ranks. 

\Fig{pt} shows that as the underlying tensor rank increases, the percentage of outliers that can be tolerated decreases for both methods. However, at a given rank, Tucker-$\LtwoE$ can often handle a greater level of corruption. At rank 25, both methods are able to obtain an accurate reconstruction for any outlier percentage no greater than 0.5. From rank 29 to 41, Tucker-$\LtwoE$ can generally handle 10-20\% more outliers. At rank 43 or 45, Tucker-$\LtwoE$ can handle 5\% more outliers. Interestingly, the advantage of Tucker-$\LtwoE$ over HoRPCA-C is notably more significant on data generated by the CP model, especially at ranks 43 and 45. This is potentially because a CP-rank of 45 is a more constrained low-rank structure than a general Tucker-rank of (45,45,45).

\subsection{Rank Misspecification and Cross Validation}\label{sec:rm}

In the previous two sections, we set the specified ranks to be equal to the true ranks for the rank-constrained formulations. In practice, however, such prior knowledge of tensor rank may not always be available. In this section, we investigate how rank-constrained formulations behave when the tensor rank is underestimated or overestimated. We also demonstrate the application of cross-validation to choose the appropriate rank for HoRPCA-C and Tucker-$\LtwoE$. We consider three scenarios: 1) the noiseless tensor has CP-rank 15; 2) the noiseless tensor has Tucker-rank (30,10,5); 3) the noiseless tensor has Tucker-rank (35,35,35). After generating the true low-rank tensor, 25\% percent of tensor entries are corrupted with outliers. The specified CP-ranks are $5,10,\dots,45$ and the specified Tucker-ranks are $(5,5,5),(10,10,10),\dots,(45,45,45)$. Notice that although in the second scenario, the noiseless tensor is not equally low-rank in every mode, we still set the specified Tucker-ranks to be equal in every mode simply to limit the number of Tucker-rank tuples that we need to consider. 
\begin{figure}[t]
  \centering
  \includegraphics[width=0.95\linewidth]{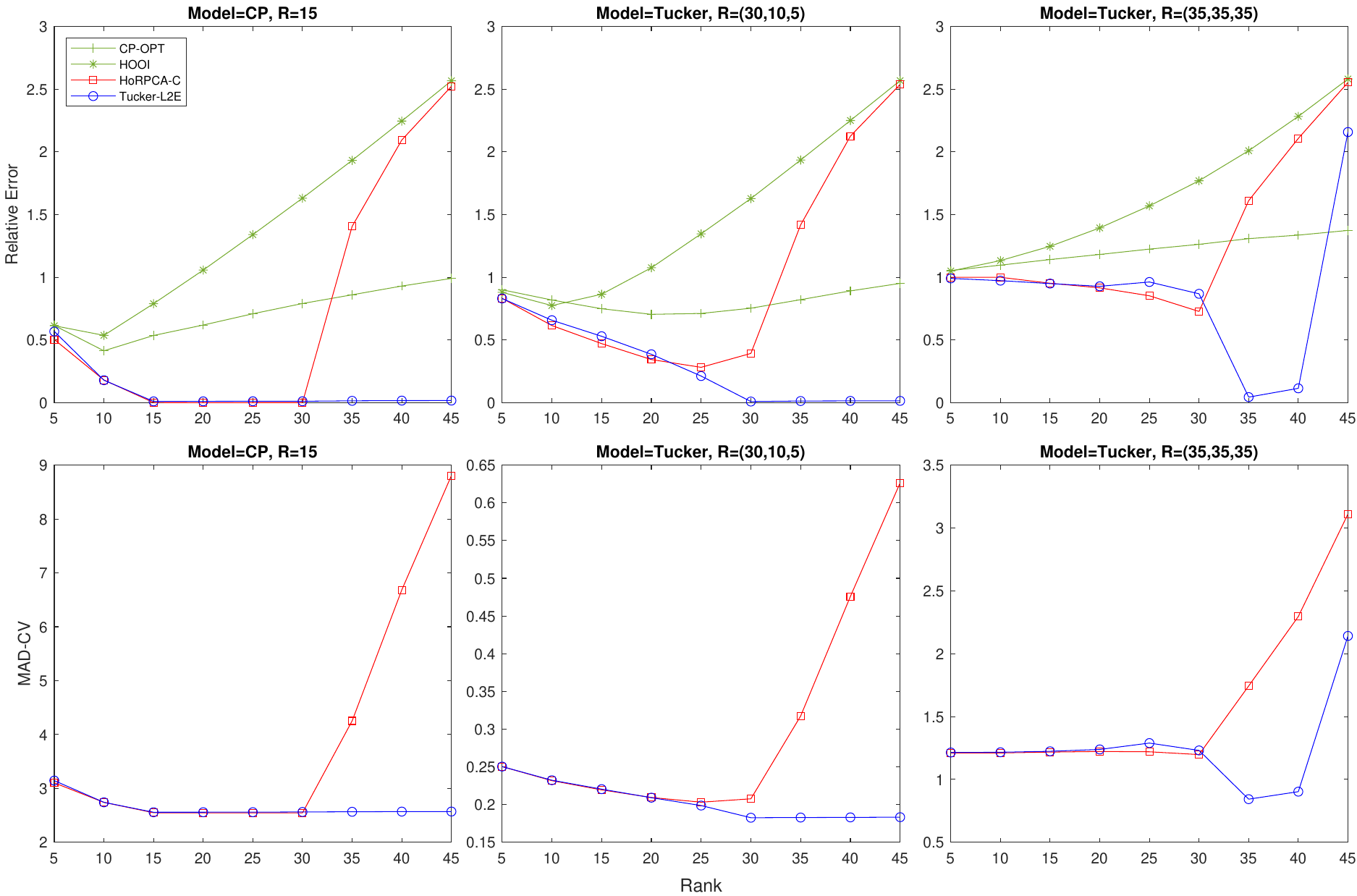}
  \caption{First row: recovery results when the rank is underestimated or overestimated. Second row: 10-fold cross-validation error for a generated tensor.}
  \label{fig:cv}
\end{figure}

The cross-validation scheme can be described as follows: the tensor entries are randomly split into 10 folds; robust tensor decomposition methods (HoRPCA-C and Tucker-$\LtwoE$) are applied to 9 out of 10 folds, treating the hold-out fold as missing data. This process is repeated for each train/test split; we use the estimated values for the hold-out fold to form a new tensor which we call the predicted tensor; cross-validation error is computed as the MAD between entries of the predicted tensor and entries of the original noisy tensor. The MAD is chosen over the more common mean squared error (MSE) to make the cross-validation error less sensitive to large residuals, which likely coincide with outlying entries.

\Fig{cv} shows that the non-robust methods (CP-OPT and HOOI) will greatly overfit to the outliers if the tensor rank is overestimated. For the first scenario ($R=15$), both HoRPCA-C and Tucker-$\LtwoE$ exhibit a certain level of overfitting resistance. Remarkably, Tucker-$\LtwoE$ remains unaffected by outliers even if the Tucker-rank is grossly overestimated to be (45,45,45). In the second scenario ($R= (30,10,5)$), unlike Tucker-$\LtwoE$, HoRPCA-C is not able to achieve perfect recovery if the Tucker-rank is specified to (30,30,30). The first two scenarios demonstrate that Tucker-$\LtwoE$ is more robust to rank overestimation than HoRPCA-C. The third scenario ($R= (35,35,35)$) is chosen to be challenging. Contrary to our expectation, the relative error and cross-validation error for Tucker-$\LtwoE$ will first increase before it decreases. It is reassuring that the cross-validation error still achieves its minimum at the true Tucker-rank. Another surprising observation is that HoRPCA-C attains its minimum relative error at Tucker-rank $(30,30,30)$ instead of the true rank $(35,35,35)$. This is likely because while at Tucker-rank $(30,30,30)$, HoRPCA-C is only capable of an approximate reconstruction, it is still better than the true Tucker-rank $(35,35,35)$ where HoRPCA-C becomes unstable.

\subsection{Varying Degrees of Missingness}\label{sec:md}
\begin{figure}[htbp]
  \centering
  \includegraphics[width=0.9\linewidth]{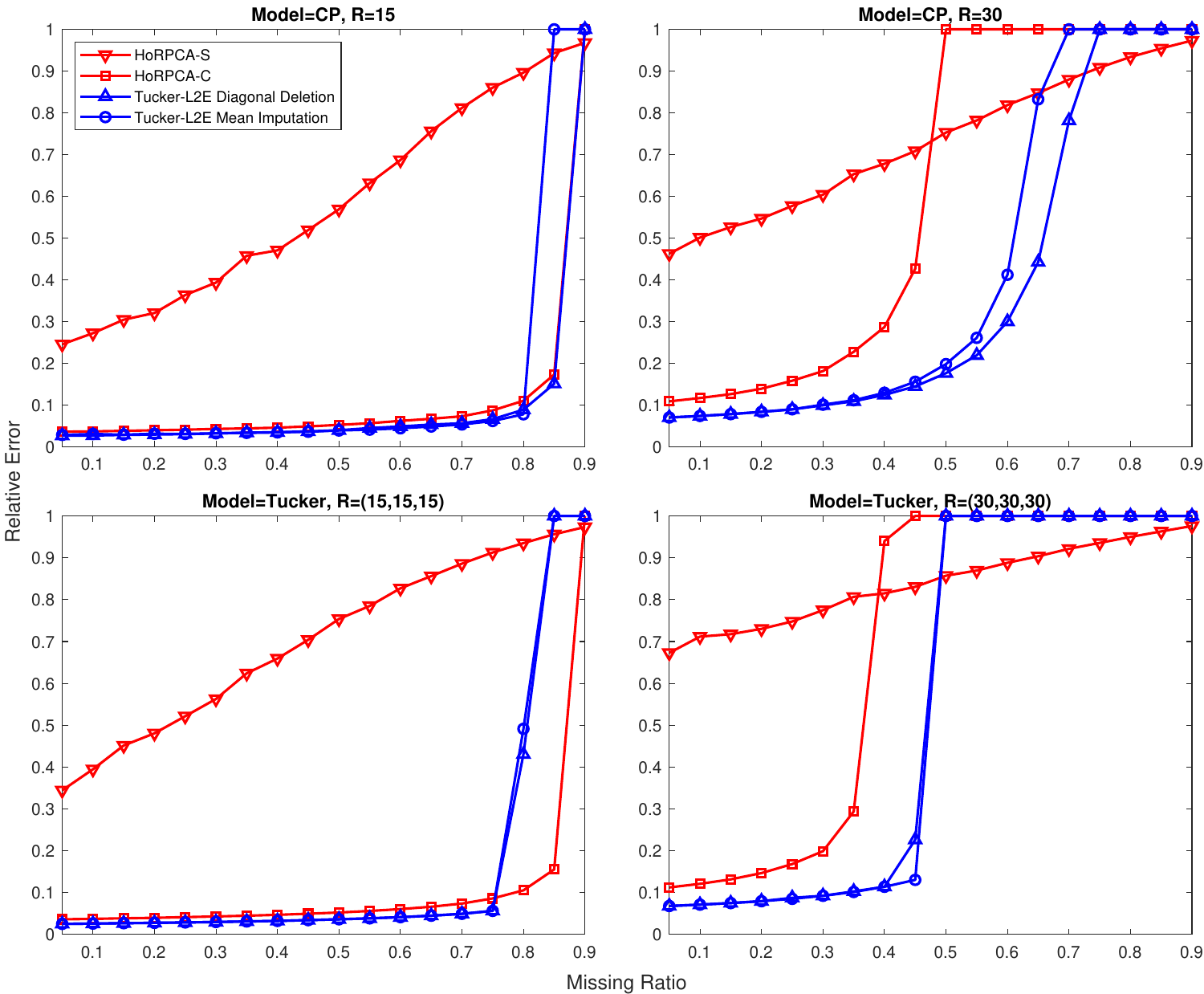}
  \caption{Recovery results under varying degrees of missingness. Outlier sparsity is set at 25\%. Data points are averaged over 20 random replicates. Average relative errors larger than 1 are capped at 1 to suit the display.}
  \label{fig:missing}
\end{figure}
In this section we investigate the recovery performance of HoRPCA-S, HoRPCA-C and Tucker-$\LtwoE$ under varying degrees of missingness. RGrad is not considered in this section since its current form does not allow missing entries. We generate $(50,50,50)$ tensors with CP-rank 15, 30 or Tucker-rank $(15,15,15)$, $(30,30,30)$. After generating the low-rank tensor, 25\% of the entries are corrupted with outliers and dense noise of relative scale 0.1 is added. Then $\rho\times 100\%$ of the tensor entries are set to be missing. The missingness is assumed to be completely random and independent from the outlier corruption. We vary the missing ratio $\rho$ from 0.05 to 0.9 in this section. We present the recovery results of both initialization methods.

In \Fig{missing}, ``Tucker-$\LtwoE$ Diagonal Deletion" refers to the recovery results using spectral initialization with diagonal deletion and ``Tucker-$\LtwoE$ Mean Imputation"  refers to the recovery results using \Alg{init1} as the initialization. We can see that the diagonal deletion procedure offers some improvement over \Alg{init1} when the underlying tensor is of low CP-rank. Tucker-$\LtwoE$ appears to be less stable than HoRPCA-C when the rank is low and the missing percentage is very high (over 80\%). This may be attributed to the fact that HoRPCA-C models missing data with an equality constraint so that the unobserved entries are still penalized for having a large magnitude, while for Tucker-$\LtwoE$ the unobserved entries are masked and unconstrained. When the rank is relatively high, we can see that Tucker-$\LtwoE$ still enjoys an empirical advantage over HoRPCA-C in terms of recovery capability.

\section{Real Data Applications}\label{sec:rda}
\subsection{Tensor Denoising on 3D fMRI Data}\label{sec:fmri}
We consider a 3D MRI dataset INCISIX from the OsiriX repository\footnote{https://www.osirix-viewer.com}, which contains 166 slices through a human brain, each having dimension $512\times512$.  The dataset was first analyzed in \cite{gandy2011tensor} from a tensor completion perspective by randomly setting voxels to be missing. We approach this dataset from a tensor denoising perspective. Following \cite{goldfarb2014robust}, we extract the first 50 slices and downsample each of them to have size $128\times128$. Therefore the noiseless tensor $\T{X}$ has dimension $128\times 128\times 50$. We then corrupt 25\% of the tensor entries with outliers/noise drawn from $\operatorname{Unif}[0,2\std(\vec(\T{X}))]$.
\begin{figure}[tbp]
  \centering
  \includegraphics[width=0.9\linewidth]{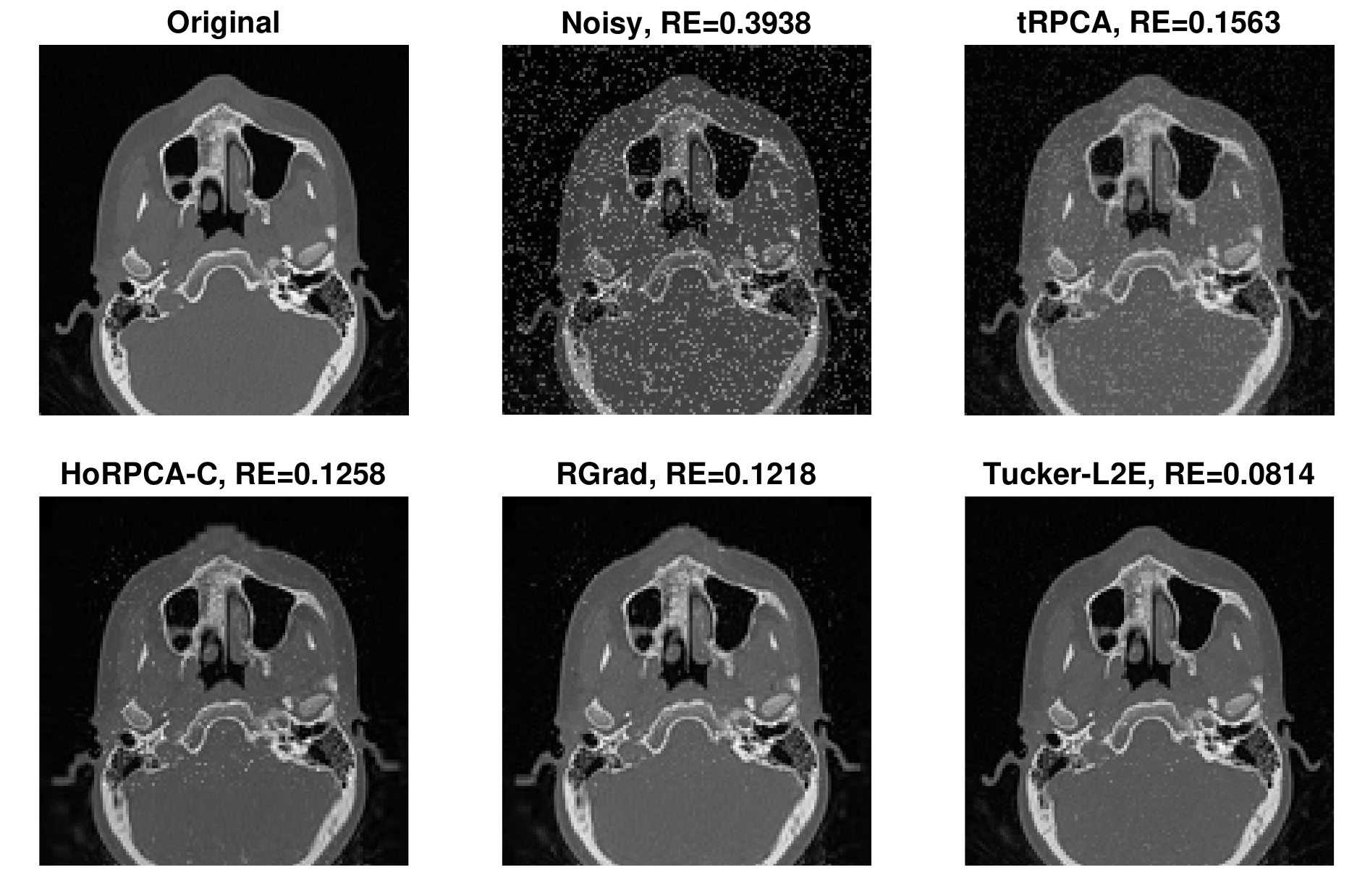}
  \caption{Recovery results for slice 30 of INCISIX dataset. The annotated relative errors are for the whole tensor instead of one slice. }
  \label{fig:fMRI}
\end{figure}
In addition to HoRPCA-C and RGrad, we consider a classic low-tubal-rank robust tensor recovery method called tensor robust principal component analysis \citep{lu2019tensor} as another baseline. Scree plots of the different matrix unfoldings revealed that the mode-3 singular values decay rapidly, which indicates that the data tensor is approximately low-rank along mode-3. Therefore, we considered  the following three Tucker-rank tuples, (64,64,10), (96,96,15) and (128,128,20), as candidate Tucker-rank tuples for HoRPCA-C, RGrad and Tucker-$\LtwoE$.  For RGrad, the proposed BIC criterion selected the parameters $R=(96,96,15)$, $\alpha=0.27$ and $\mu_0=1$. For HoRPCA-C and Tucker-$\LtwoE$, we use a hold-out validation approach to identify the best Tucker-rank tuple. We randomly sample 10\% of the entries as the validation set and use the remaining 90\% of entries to impute the missing 10\% of entries. Then we can use the MAD between the imputed values and the actual values to determine the appropriate Tucker-rank. From \Tab{holdout}, we see that the best performing Tucker-rank for HoRPCA-C is $(96,96,15)$, while for Tucker-$\LtwoE$ it is $(128,128,20)$. We then reapply HoRPCA-C and Tucker-$\LtwoE$ with the selected ranks to the fully observed tensor.
\begin{table}[htbp]
\centering
\begin{tabular}{cccc}
\hline
Tucker-rank                  & $(64,64,10)$ & $(96,96,15)$    & $(128,128,20)$  \\ \hline
HoRPCA-C hold-out MAD         & 223.16       & \textbf{216.85} & 276.60          \\
Tucker-$\LtwoE$ hold-out MAD & 236.51       & 211.12          & \textbf{206.60} \\ \hline
\end{tabular}
\caption{Hold-out MAD for HoRPCA-C and Tucker-$\LtwoE$ at different Tucker-ranks. }
\label{tab:holdout}
\end{table}

We visualize the recovery results of tRPCA, HoRPCA-C, RGrad and Tucker-$\LtwoE$ in \Fig{fMRI}. The advantage of Tucker-$\LtwoE$ over HoRPCA-C and RGrad is that at rank (128,128,20), HoRPCA-C and RGrad overfit to the sparse noise while Tucker-$\LtwoE$ remains largely unaffected, in line with our observations  in \Sec{ne}. By remaining stable at a larger rank, Tucker-$\LtwoE$ is able to capture more structural information, resulting in the smallest relative error and the best perceptual quality.
\subsection{PARAFAC Analysis of Fluorescence Data}
Parallel Factor analysis (PARAFAC) is a widely used tool in Chemometrics for decomposing fluorescence excitation-emission matrices (EEMs) into their underlying chemical components \citep{murphy2013fluorescence}. The CP model is particularly suitable for the analysis of EEMs since this type of data mostly conforms to a trilinear structure due to Beer's law, which states that absorbance is the product of molar concentration, molar absorption coefficient, and optical path length. Certain regions of the fluorescence landscape, however, may be corrupted by Raman and Rayleigh scattering.  Therefore, EEM data is a natural candidate for  the low CP-rank tensor plus sparse outliers model. 

\begin{figure}[htbp]
  \centering
  \includegraphics[width=0.7\linewidth]{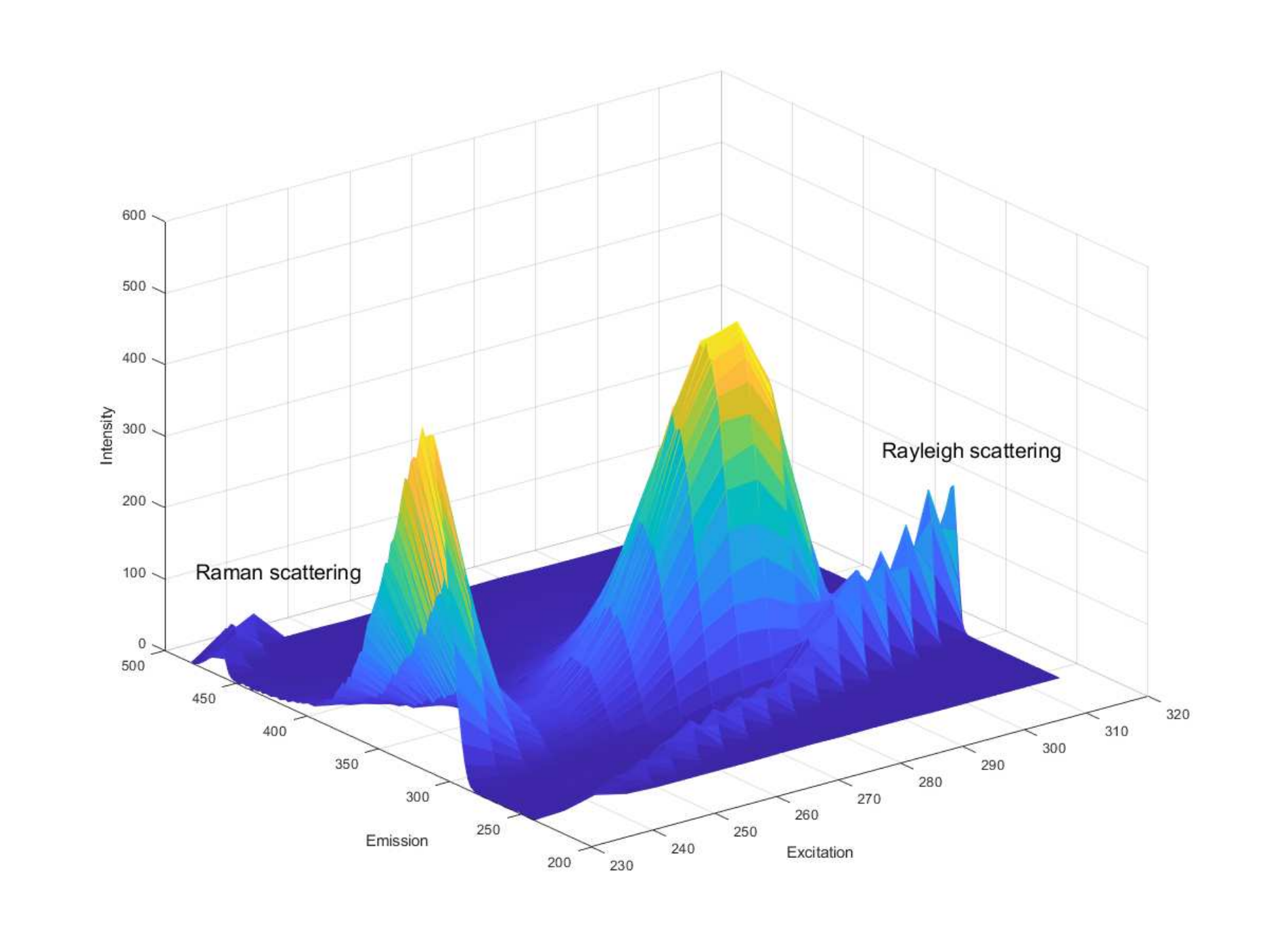}
  \caption{Truncated fluorescence landscape of sample 1 in the Dorrit data. Intensity peaks caused by Raman and Rayleigh scattering can be observed on the top left and the bottom right.}
  \label{fig:landscape}
\end{figure}

We consider a standard EEM dataset, the Dorrit data, originally introduced in \cite{dorrit}. We use a preprocessed version\footnote{http://www.models.life.ku.dk/dorrit} \citep{riu2003jack} of the Dorrit data, which consists of 27 mixed samples containing different concentrations of hydroquinone, tryptophan, phenylalanine, and dopa. Each sample has 121 emission wavelengths (241-481 nm) and 24 excitation wavelengths (200-315 nm). Following  \cite{riu2003jack} and \cite{goldfarb2014robust}, we exclude samples 2, 3, 5, and 10 as well as data corresponding to  excitation wavelengths from 200 nm to 225 nm since this portion of the data is believed to be noisy for reasons other than scattering and amounts to slice-wise corruption, which greatly affect the global properties of the data. Therefore the tensor data to be analyzed have dimension $23\times 121 \times 18$. The truncated fluorescence landscape of sample 1 is visualized in \Fig{landscape}. We set the CP-rank to 4 and the Tucker-rank to $(4,4,4)$ since we have prior knowledge that there are 4 pure compounds in the samples. 
\begin{figure}[htbp]
  \centering
  \includegraphics[width=\linewidth]{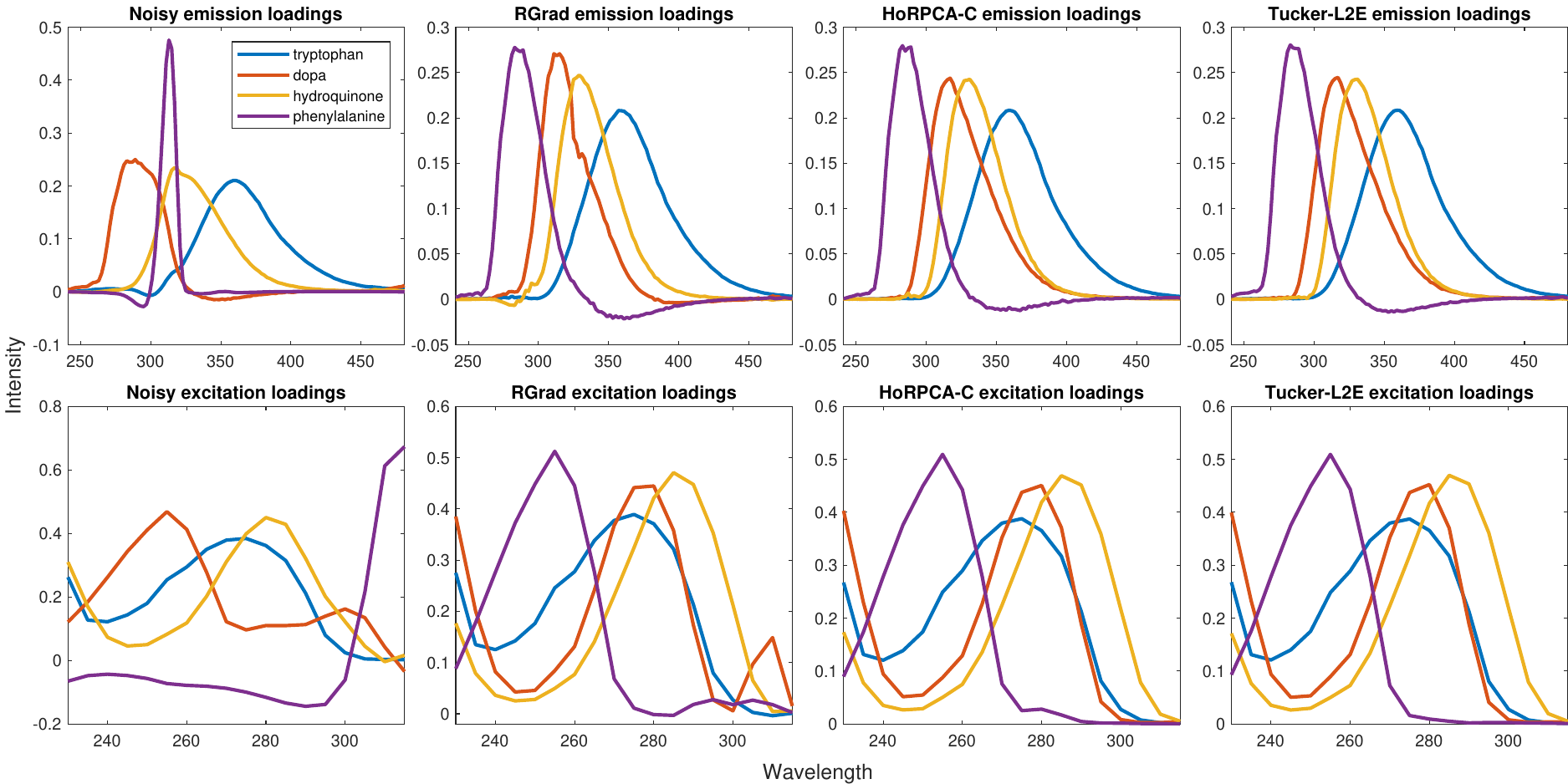}
  \caption{Reconstructed mode-2 and mode-3 CP factors for Dorrit data with scattering only.}
  \label{fig:dorrit}
\end{figure}

\Fig{dorrit} displays the recovered emission/excitation loadings  (mode-2 and mode-3 CP factors) produced by CP-OPT, RGrad, HoRPCA-C, and Tucker-$\LtwoE$. The last three methods are in fact applied to denoise the tensors. The CP factors are then extracted by applying CP-OPT to the reconstructed tensors.  We assign the CP factors to the 4 analytes based on proximity to the pure component  emission/excitation profiles presented in \cite{dorrit}.  For this dataset, with rank fixed at $(4,4,4)$, BIC suggests $\alpha=0.11$ and $\mu_0=5$ for RGrad. This has an interesting implication that there are approximately 11\% of entries that are affected by scattering. The emission/excitation loadings produced by HoRPCA-C and Tucker-$\LtwoE$ appear more similar to the pure component profile than the ones produced by RGrad. Although the difference between HoRPCA-C and Tucker-$\LtwoE$ is minuscule, the emission/excitation loadings of phenylalanine (purple lines in \Fig{dorrit}) produced by Tucker-$\LtwoE$ appear to be slightly more regular than those produced by HoRPCA-C.
\subsection{Feature Extraction for Classification}\label{sec:fe}
Tucker decompositions are useful for extracting features from high-dimensional multi-way datasets for classification. The extracted features can then be used as input to standard classifiers such as \textit{k-nearest-neighbors} (k-NN) or \textit{support vector machines} (SVM). In this section we adopt the feature extraction framework based on Tucker decomposition originally presented in \cite{phan2010tensor}. \cite{chachlakis2019l1} demonstrated that if the dataset is corrupted with sparse noise, strategic dimensionality reduction by a Tucker decomposition can reduce the impact of noise and leads to improved classification accuracy. 
Below we briefly describe the feature extraction framework. Suppose that we have $K_1$ tensor-valued training samples of size $I_1\times I_2\times\dots\times I_N$, which can be classified into $C$ categories. We concatenate training samples across mode $N+1$ to obtain  $\T{X}_1\in\Real^{I_1\times I_2\times\dots\times I_N\times K_1}$, which we call the ``training tensor." Then $\T{X}_1$ is  Tucker decomposed with rank $(d_1,d_2,\dots,d_N,K_1)$ to obtain the factor matrices $\M{U}_n\in \Real^{I_n\times d_n}$ for $n \in [N]$. We compress the training samples as follows:
\begin{eqnarray*}
\T{Z}_1 & = & \T{X}_1\times_1\M{U}_1\Tra\times_2\M{U}_2\Tra\times_3\dots\times_N\M{U}_N\Tra \amp \in\amp \Real^{d_1\times d_2 \times\dots \times d_N\times K_1}.
\end{eqnarray*}
Then  $\T{Z}_1$ is matricized in mode $N+1$ to become a data matrix $\M{Z}_1$ of size $K_1 \times \prod_{n=1}^N d_n$ with labeled rows. We similarly concatenate testing samples to obtain the ``testing tensor" $\T{X}_2\in \Real^{d_1\times d_2 \times\dots \times d_N\times K_2}$, which we compress to obtain
\begin{eqnarray*}
\T{Z}_2 & = & \T{X}_2\times_1\M{U}_1\Tra\times_2\M{U}_2\Tra\times_3\dots\times_N\M{U}_N\Tra \amp \in \amp \Real^{d_1\times d_2 \times\dots \times d_N\times K_2},
\end{eqnarray*}
which is then matricized in mode $N+1$ to become $\M{Z}_2\in \Real^{K_2 \times \prod_{n=1}^N d_n}$. A suitable classifier is then trained on $\M{Z}_1$ and tested on $\M{Z}_2$.  
\begin{figure}[tbp]
     \centering
     \begin{subfigure}[b]{0.45\textwidth}
         \centering
         \includegraphics[width=\textwidth]{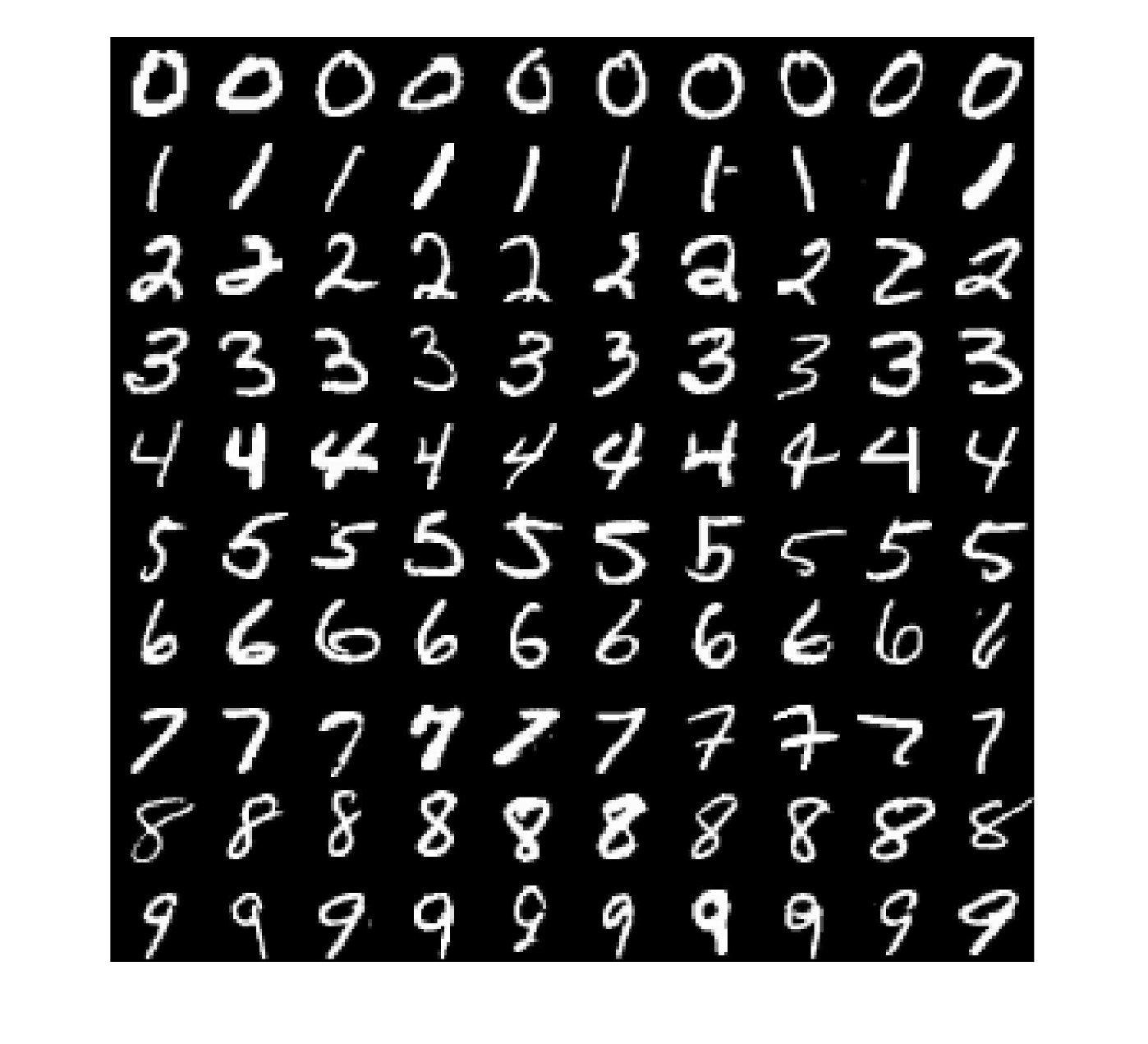}
         \caption{Randomly sampled images from the MNIST dataset.}
     \end{subfigure}
     \hfill
     \begin{subfigure}[b]{0.45\textwidth}
         \centering
         \includegraphics[width=\textwidth]{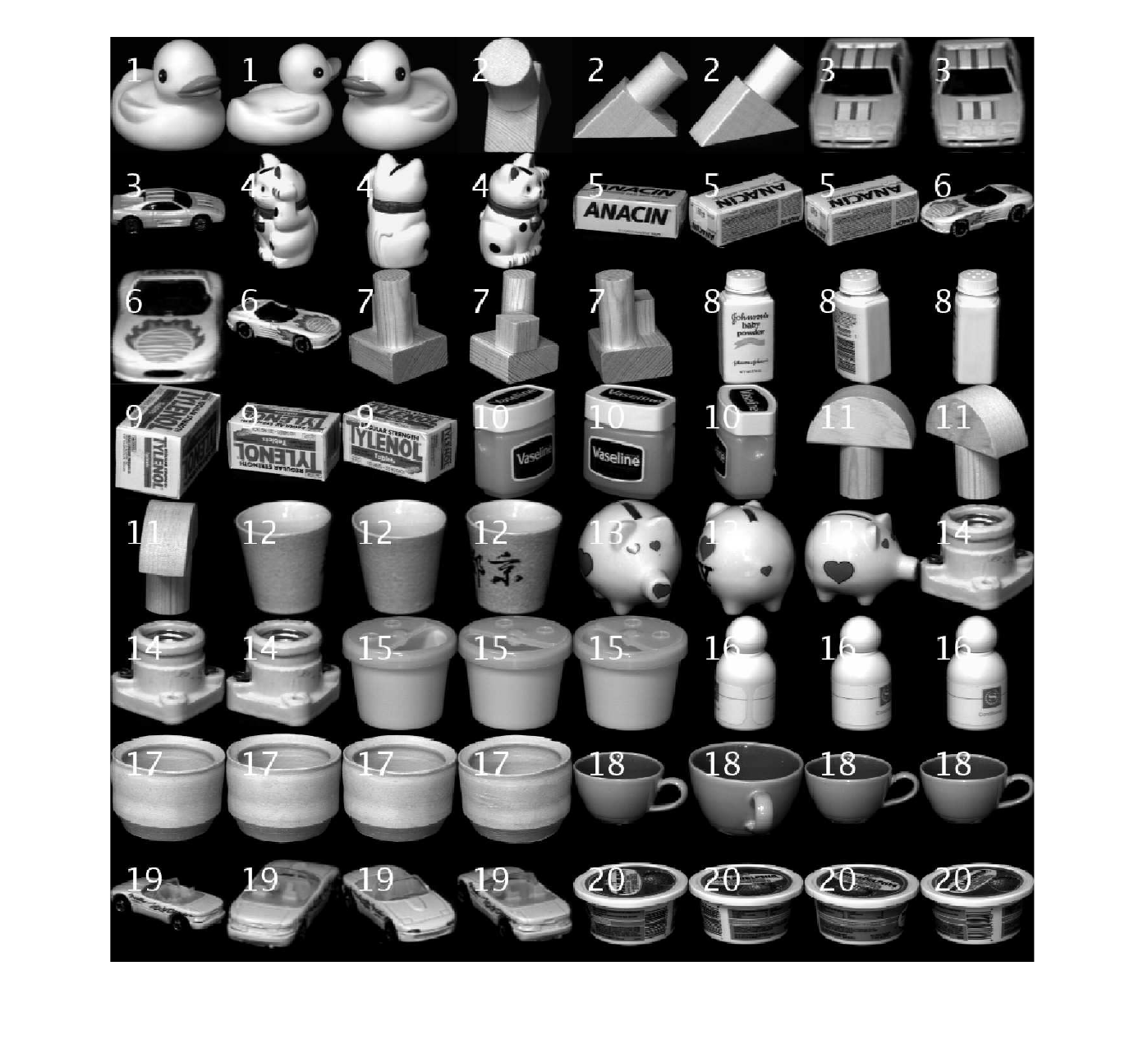}
         \caption{Randomly sampled images from the COIL-20 dataset.}
     \end{subfigure}
     \caption{Visualizations of the MNIST dataset and the COIL-20 dataset.}
  \label{fig:mnist_coil}
\end{figure}
\begin{figure}[htbp]
  \centering
  \includegraphics[width=\linewidth]{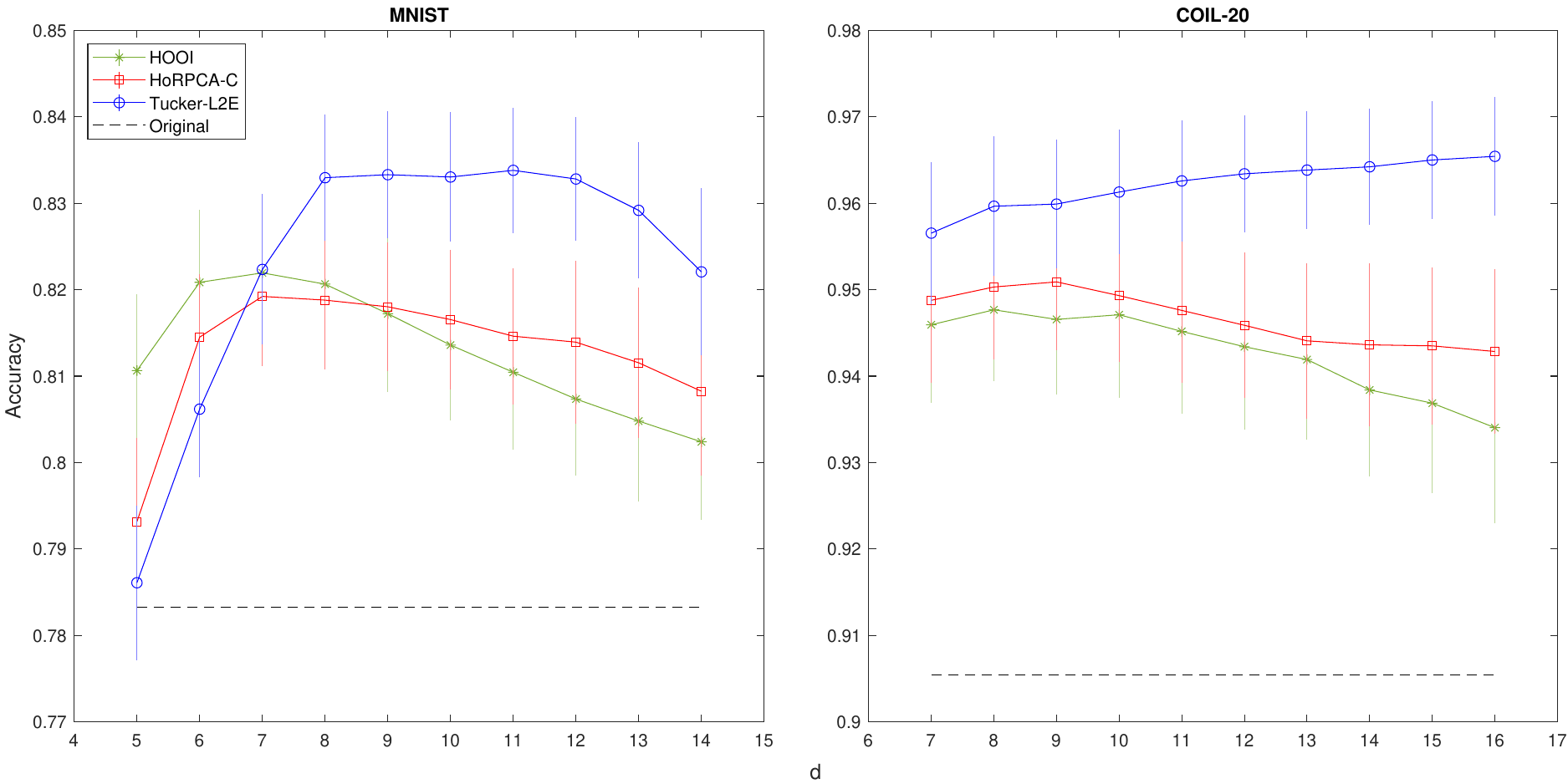}
  \caption{Average testing accuracy across 50 random realizations of training/testing sets. Error bars denote $\pm 1$ standard errors. Gray line shows the testing accuracy of \textit{nearest-neighbor} without applying Tucker decomposition for dimensionality reduction.}
  \label{fig:classification}
\end{figure}
We apply the above feature extraction framework, with an added aspect of robustness, to image classification. We consider two classic image classification datasets, namely MNIST \citep{deng2012mnist} and COIL-20 \citep{nene1996columbia}, which are also studied in \cite{phan2010tensor}. MNIST consists of $28\times 28$ images of hand-written digits. COIL-20 consists of $128\times 128$ images for 20 different objects with each object having 72 images. \Fig{mnist_coil} depicts samples of images from both datasets.

As a preprocessing step, we remove the 4-pixel padding for MNIST since those regions contain no information. Images from COIL-20 are downsampled to $32\times 32$ to speed up computation. To construct a training sample, we randomly sample 50 images for each digit from MNIST and 20 images for each object from COIL-20. The training images are then corrupted with salt-and-pepper noise added to a randomly selected sample of 25\% of the pixels. Thus for MNIST, $\T{X}_1$ has dimension $20\times 20\times 500$ while for COIL-20, $\T{X}_1$ has dimension $32\times 32\times 400$. We set $d_1=d_2=d$ so that the total number of features is $d^2$. We randomly sample another 500 images for each digit and use the remaining 52 images for each object as the testing points. The classifier of choice is \textit{nearest-neighbor}. We repeat the described procedure on 50 different random realizations of training and testing sets. The Tucker decomposition methods considered here are HOOI, HoRPCA-C and Tucker-$\LtwoE$. In particular, HoRPCA-C is only used to approximate the training tensor. The factor matrices are obtained by applying HOOI to the output tensor of HoRPCA-C.

\Fig{classification} highlights that by applying Tucker decomposition for feature extraction and dimensionality reduction, all methods achieve a substantial gain in accuracy compared with directly using the corrupted training images. As $d$ increases, initially the accuracy of all Tucker decomposition methods will increase due to being able to create more meaningful features. However, eventually the feature extraction framework will overfit to the sparse noise and the accuracy starts decreasing. The testing accuracy on MNIST is generally much lower than on COIL-20 despite having fewer categories and more training images per category. This suggests that MNIST is a more challenging dataset to classify. Tucker-$\LtwoE$ again exhibits greater stability in high-rank scenarios, especially in the case of COIL-20 with its accuracy steadily increasing throughout the range of $d$ that we investigated. When $d$ is small, Tucker-$\LtwoE$ may not be advantageous. However, as the number of features increases, the best attainable accuracy of Tucker-$\LtwoE$ outperforms that of HOOI and HoRPCA-C.

\section{Conclusion}
This paper describes a new formulation of the robust Tucker decomposition problem based on the $\Ltwo$ criterion, Tucker-$\LtwoE$. We present two initialization strategies and a solution algorithm based on L-BFGS-B. Numerical experiments and real data applications demonstrate that Tucker-$\LtwoE$ exhibits stronger recovery capability in challenging high-rank scenarios compared with existing alternatives. This empirical property is useful since real-world tensors are often nearly low-rank instead of perfectly low-rank. By remaining stable at a higher rank, Tucker-$\LtwoE$ is able to provide a more expressive reconstruction of the underlying low-rank tensor in the presence of sparse perturbations.

In this article we used an off-the-self local optimization algorithm L-BFGS-B as the main computational tool. A projected-gradient type algorithm will likely have a smaller memory footprint, which presents an interesting venue for future work. We also note that the proposed robust tensor recovery paradigm can be adapted to other formats of  low-rank tensor recovery, for example the low-tubal-rank format and the tensor-train format, with suitably designed computation algorithms.  
\begin{center}
    {\large \textbf{Supplementary Materials}}
\end{center}
\begin{description}
\item[Supplement:] A pdf file that contains derivation of gradient, an alternative initialization  strategy named spectral initialization with diagonal deletion, proof of \Thm{well},  details of parameter choices, and a run time comparison with the baseline methods. 
\item[Software:]  Matlab code of the described method, along with scripts to reproduce some of the figures in \Sec{ne} and \Sec{rda}. 
\end{description}
\newpage
\begin{center}
    {\large \textbf{Acknowledgement}}
\end{center}
We are grateful to the associate editor, editor, and anonymous referee for their valuable comments and suggestions, which greatly improved the presentation of this paper. We thank Haixu Ma and Xiaoqian Liu for their assistance in testing the software.
\begin{center}
    {\large \textbf{Funding}}
\end{center}
This research was partially funded by grants from the National Institute of General Medical Sciences (R01GM135928: EC, R01GM126550: YL) and the National Science Foundation (DMS-2201136: EC, DMS-2100729: YL).
\begin{center}
    {\large \textbf{Disclosure Statement}}
\end{center}
The authors report there are no competing interests to declare.

\bibliographystyle{asa}
\bibliography{refs}
\end{document}